\documentclass[structabstract]{aa} 

\usepackage{amssymb}
\usepackage{amsmath}
\usepackage{txfonts}
\usepackage{xspace}
\usepackage{graphicx}
\usepackage{epsfig,epsf}
\usepackage{natbib}
\usepackage{color}
\usepackage[colorlinks=true, linkcolor=blue, citecolor=blue, urlcolor=blue, breaklinks=true]{hyperref}
\definecolor{red}{rgb}{1.00,0.00,0.00}
\bibpunct{(}{)}{;}{a}{}{,} 
\begin{document}

\title{\textit{HERschel}\thanks{\textit{Herschel} is an ESA space observatory with science 
instruments provided by European-led Principal Investigator consortia and with important 
participation from NASA.} Observations of Edge-on Spirals (\textit{HER}OES).}  
\subtitle{I: Far-infrared morphology and dust mass determination}

\author{
J.~Verstappen\inst{\ref{inst-UGent}},
J.~Fritz\inst{\ref{inst-UGent}},
M.~Baes\inst{\ref{inst-UGent}},  
M.~W.~L.~Smith\inst{\ref{inst-CardU}}, 
F.~Allaert\inst{\ref{inst-UGent}},
S.~Bianchi\inst{\ref{inst-Arcetri}}, 
J.~A.~D.~L.~Blommaert\inst{\ref{inst-KUL},\ref{inst-VUB}},
G.~De~Geyter\inst{\ref{inst-UGent}}
I.~De~Looze\inst{\ref{inst-UGent}}, 
G.~Gentile\inst{\ref{inst-UGent},\ref{inst-VUB}},
K.~D.~Gordon\inst{\ref{inst-STScI},\ref{inst-UGent}}
B.~W.~Holwerda\inst{\ref{inst-ESTEC}},
S.~Viaene\inst{\ref{inst-UGent}}
\and
E. M. Xilouris\inst{\ref{inst-Athens}}
}  

\institute{ 
  Sterrenkundig Observatorium, Universiteit Gent, Krijgslaan
  281, B-9000 Gent, Belgium\label{inst-UGent}
  \and 
  School of Physics and Astronomy, Cardiff University, Queens Buildings, The Parade,
  Cardiff CF24 3AA, UK\label{inst-CardU}
  \and 
  INAF - Osservatorio Astrofisico di Arcetri, Largo E. Fermi 5, 50125, Florence,
  Italy\label{inst-Arcetri}
  \and 
  Instituut voor Sterrenkunde, K. U. Leuven, Celestijnenlaan 200D, B-3000 Leuven,
  Belgium\label{inst-KUL}
  \and
  Vakgroep Fysica en Sterrenkunde, Vrije Universiteit Brussel, Pleinlaan 2, 1050 Brussels, Belgium\label{inst-VUB}
  \and
  Space Telescope Science Institute, 3700 San Martin Drive, Baltimore,
  MD 21218, USA\label{inst-STScI}
  \and 
  European Space Agency, ESTEC, Keplerlaan 1, 2200 AG, Noordwijk, the
  Netherlands\label{inst-ESTEC}
  \and 
  Institute for Astronomy, Astrophysics, Space Applications \& 
  Remote Sensing, National Observatory of
  Athens, P. Penteli 15236 Athens, Greece\label{inst-Athens}
}

\date{\today}

\abstract
{Edge-on spiral galaxies with prominent dust lanes provide us with an
  excellent opportunity to study the distribution and properties of
  the dust within them. The \textit{HER}OES project was set up
  to observe a sample of seven large edge-on galaxies across various
  wavelengths for this investigation.}
{Within this first paper, we present the \textit{Herschel}
  observations and perform a qualitative and quantitative analysis on
  them, and we derive some global properties of the far infrared and
  submillimetre emission.}
{We determine horizontal and vertical profiles from the
  \textit{Herschel} observations of the galaxies in the sample
  and describe the morphology. Modified black-body fits to the global
  fluxes, measured using aperture photometry, result in dust
  temperatures and dust masses.  The latter values are compared to
  those that are derived from radiative transfer models taken from the
  literature.}
{On the whole, our \textit{Herschel} flux measurements agree well with
  archival values. We find that the exponential horizontal dust
  distribution model often used in the literature generally provides a
  good description of the observed horizontal profiles.
    Three out of the seven galaxies show signatures of extended
    vertical emission at 100 and 160 $\mu$m at the 5$\sigma$ level,
    but in two of these it is probably due to deviations from an
    exactly edge-on orientation. Only for NGC\,4013, a galaxy in which
    vertically extended dust has already been detected in optical
    images, we can detect vertically extended dust, and the derived
    scaleheight agrees with the value estimated through
    radiative transfer modelling.  Our analysis hints at a
  correlation between the dust scaleheight and its degree of
  clumpiness, which we infer from the difference between the dust
  masses as calculated from modelling of optical data and from fitting
  the spectral energy distribution of \textit{Herschel} datapoints.}
{}

\keywords{Galaxies: structure -- Infrared: ISM -- Infrared: galaxies
  -- Submillimeter: ISM -- Submillimeter: galaxies -- Dust,
  extinction}

\titlerunning{\textit{HER}OES I}
\authorrunning{J. Verstappen et al.}

\maketitle

\section{Introduction}
\label{Introduction.sec}

In the past three decades, interstellar dust has changed from being
mainly a nuisance that hampered any correct interpretation of optical
data to a fascinating and important component of the interstellar
medium in galaxies. Indeed, dust grains not only absorb and scatter
light in the optical and ultraviolet (UV), but they also help
regulate the physics and chemistry of the interstellar medium and play
a crucial role in star and planet formation. Unfortunately, it is
observationally hard to determine the amount, spatial distribution and
physical properties of the interstellar dust in galaxies. The most
straightforward way to trace the dust in galaxies is by looking at
far-infrared (FIR) and submillimetre (sub-mm) wavelengths, where the
emission by cold dust grains dominates the spectral energy
distribution (SED). Until recently, the sensitivity, spatial
resolution and wavelength coverage of the available FIR/sub-mm
instrumentation has been rather limited, and most of our knowledge was
often restricted to studying the global SED of galaxies
\citep[e.g.][]{1987ApJ...320..238S, 2000MNRAS.315..115D,
  2001ApJ...557...39P, 2002ApJS..139...37T, 2007ApJ...663..866D}. This
situation has changed substantially with the launch of the
\textit{Herschel Space Observatory}
\citep{2010A&A...518L...1P}, which has a much improved spatial
resolution and wider wavelength coverage than any of its
predecessors. With \textit{Herschel}, we can now take spatially
resolved FIR/sub-mm images to map the distribution of cool dust (the
bulk of the dusty ISM) in nearby galaxies
\citep[e.g.][]{2010A&A...518L..51S, 2012ApJ...756...40S,
  2010A&A...518L..65B, 2012MNRAS.419.1833B, 2010A&A...518L..72P,
  2011PASP..123.1347K, 2012MNRAS.421.2917F, 2012A&A...546A..34F,
  2012A&A...539A.145B, 2012A&A...543A..74X, 2012ApJ...756..138A,
  2012ApJ...755..165M, 2012MNRAS.423.2359D}.

Edge-on spiral galaxies offer an interesting perspective for studying
the dust properties and distribution in spiral discs. The dust in
edge-on spiral galaxies often shows up as prominent dust lanes in
optical images, which makes this class of galaxies among the only
systems where dust can easily be studied both in extinction and in
emission. They are also the only systems where the vertical
distribution of the dust can be studied. Due to line-of-sight
projection, edge-on spirals also allow us to map the horizontal
distribution of dust in detail, since the increased surface brightness
enables us to trace the dust extinction and emission to large radial
distances from the centre. Cold dust has been located in substantial
quantities at large galactocentric radii for a small number of edge-on
spirals using FIR and sub-mm observations before
\citep[e.g.][]{1998ApJ...507L.125A, 2003MNRAS.344..105D,
  2003A&A...410L..21P}, but the limited sensitivity of the FIR
instruments of the previous generation was a strong constraint on
attempts to study the horizontal dust distribution for a larger sample
of edge-on spiral galaxies.

The strongest constraints on the distribution and properties of dust
in spiral galaxies can be achieved by studying the dust energy balance
through a self-consistent treatment of extinction and thermal
emission. Due to their special orientation with respect to us, edge-on
spiral galaxies are the ideal targets for such energy balance studies.
Several edge-on spiral galaxies have undergone dust distribution
modelling by fitting realistic radiative transfer models to optical
images \citep{1987ApJ...317..637K, 1997A&A...325..135X,
  1998A&A...331..894X, 1999A&A...344..868X, 2004A&A...425..109A,
  2007A&A...471..765B, 2010A&A...518L..39B,
  2012MNRAS.427.2797D}. These studies suggest that, at least for
massive galaxies, the dust is distributed in a disc that is vertically
thinner but horizontally more extended (by some 50\% on average)
compared to the stellar disc. The face-on optical depths found are 
typically smaller than one at optical wavelengths, indicating that the 
entire galaxy disc would be almost transparent when seen 
face-on. This result is somewhat in contradiction to the spiral disc 
transparency measurements using either the number of background 
galaxies \citep{2005AJ....129.1381H, 2007AJ....134.1655H} or overlapping 
galaxy pairs \citep{2000ApJ...545..171D, 2001AJ....121.1442K, 2001AJ....122.1369K, 
2009AJ....137.3000H, 2013AN....334..268H}. These studies find that spiral 
arms are opaque and discs gradually become optically thick towards the 
galaxy's centre.

The earlier results \citep{2000ApJ...545..171D, 2005A&A...444..109H,
 2007AJ....134.2385H} however 
sampled the disc in a large physical aperture ($>1$ kpc) which introduces a bias against 
transparent regions, as they were often placed by eye or include a spiral 
arm. Subsequent studies used the high resolution of HST to map the 
transparency of a large section of the disc. When the spatial resolution is
sampling below the size of a typical molecular cloud, the distribution of the disc 
transparency becomes an exponential one with a 0.3--0.5 drop-off 
depending on the Hubble Type and mass of the galaxy \citep{2009AJ....137.3000H}. 
With the inclusion of a size distribution of ISM clouds in SED models 
and more measurements of the distribution of disc transparency using occulting 
galaxy pairs, the two approaches are set to converge on a physical model 
of light transport in spiral discs.


A quantitative comparison of these spiral galaxy radiative transfer
models based on optical extinction with FIR/sub-mm emission
observations leads to an interesting discrepancy: the rather optically
thin dust discs determined from the optical modelling absorb about 10\%
of the stellar radiation, whereas the FIR/sub-mm observations of normal
spiral galaxies indicate they typically reprocess about 30\% of the UV
and optical radiation \citep{2002MNRAS.335L..41P,
  2012MNRAS.419.3505D}. This dust energy balance problem is
particularly evident when studying individual edge-on spiral galaxies:
self-consistent radiative transfer models which successfully explain
the optical extinction, predict FIR/sub-mm fluxes that underestimate
the observed values by a factor of about three
\citep{2000A&A...362..138P, 2001A&A...372..775M, 2004A&A...425..109A,
  2005A&A...437..447D, 2010A&A...518L..39B, 2012MNRAS.427.2797D}. To
solve this energy budget problem, two widely different scenarios have
been suggested.  One straightforward solution proposed is that the
  FIR/sub-mm dust emissivity has been underestimated significantly
  \citep{2004A&A...425..109A, 2005A&A...437..447D}. The other scenario
  seeks the solution in the geometrical distribution of stars and
  dust, which is impossible to disentangle precisely in edge-on
  galaxies. If a sizeable fraction of the FIR/sub-mm emission arises
  from dust having a negligible influence on the extinction of the
  bulk of the starlight, e.g.\ because it is locked up in compact
  clumps, it can produce relatively more FIR/sub-mm emission compared
  to a galaxy with stars and dust smoothly mixed
  \citep{2000A&A...362..138P, 2001A&A...372..775M,
    2008A&A...490..461B}.

As indicated earlier, \textit{Herschel} offers the possibility
to study the dust emission from spiral galaxies in more detail than
ever before. The combination of the sensitivity and the wavelength
coverage of the Photodetector Array Camera and Spectrometer
\citep[PACS,][]{2010A&A...518L...2P} and Spectral and Photometric
Imaging Receiver \citep[SPIRE,][]{2010A&A...518L...3G} instruments,
which together cover the 70 to 500~$\mu$m\ wavelength region where the
emission from cold dust dominates, enables us to make reliable
estimates of the total thermal emission of the interstellar dust. The
increase in spatial resolution compared to previous FIR
instrumentation is a huge advantage: while energy balance studies in
the past almost exclusively relied on integrated spectral energy
distributions, \textit{Herschel} observations allow us to trace
both the vertical and horizontal distribution of FIR/sub-mm emission in
detail and make the comparison with spatially resolved radiative
transfer model predictions. Studies exploiting PACS and SPIRE
observations of edge-on galaxies have already demonstrated that a
further leap forward in the comprehension of dust distribution and
characteristics is now possible \citep{2010A&A...518L..39B,
  2011A&A...531L..11B, 2012A&A...541L...5H, 2012MNRAS.427.2797D,
  2012MNRAS.419..895D}.

This paper is the first in a series devoted to the \textit{HERschel}
Observations of Edge-on Spirals (\textit{HER}OES)
project. The goal of the \textit{HER}OES project is to make a
detailed study of the amount, spatial distribution and properties of
the interstellar dust in a sample of seven large, edge-on spiral
galaxies, and to link this to the distribution and properties of
stars, interstellar gas and dark matter. This project builds strongly
on new \textit{Herschel} observations, which are crucial for a
solid determination of the distribution of cold interstellar
dust. However, we have also set up a multi-wavelength observational
campaign to map the different components in these systems, including
optical, near-infrared, H{\sc{i}} and CO observations. Moreover, we
will analyse and interpret the observational data using
state-of-the-art radiative transfer simulations.

In this first \textit{HER}OES paper, we concentrate on a presentation
and a qualitative and quantitative analysis of the \textit{Herschel}
data.  This work will be followed up by a number of
  papers which will focus on different aspects of this rich data
  set. In Paper II we will exploit the high spatial resolution of the
  {\it Herschel} images in conjunction with \textit{Spitzer} and
  \textit{WISE} data to derive the dust properties as a function of
  position, by adopting a pixel--by--pixel SED fitting approach in a
  similar fashion as done by \citet{2010A&A...518L..51S,
    2012ApJ...756...40S}.  In Paper III we will present existing and
  new optical and NIR images of the seven \textit{HER}OES galaxies,
  and we will fit detailed radiative transfer models to these
  images. This study will take advantage of FitSKIRT
  \citep{2013A&A...550A..74D}, a fitting tool built around the SKIRT Monte
  Carlo radiative transfer code \citep{2003MNRAS.343.1081B,
    2011ApJS..196...22B} designed to fully automatically recover the
  structural properties of dust and stars with a particular focus on
  edge-on systems. These models will be used to self-consistently
  predict the spatially resolved FIR/sub-mm emission from each of the
  galaxies in our sample, extending previous attempts to panchromatic
  models over the entire UV/optical/NIR range. In Paper IV we will
  combine the \textit{Herschel} data with new and archival H{\sc{i}}
  and CO data in order to study the spatially resolved gas-to-dust
  ratio in edge-on spiral galaxies out to large radii.
 
The outline of the paper is as follows: in
Section~{\ref{Observations.sec}} we describe the sample selection and
give details on the \textit{Herschel} observations and data
reduction. In Section~{\ref{GlobalFluxes.sec}} we derive the global
fluxes for the galaxies in our sample at the PACS and SPIRE
wavelengths, and we compare the values with archival fluxes at
comparable wavelengths obtained with \textit{IRAS},
\textit{ISO}, \textit{Akari} and
\textit{Planck}. In Section~{\ref{Morphology.sec}} we discuss
the FIR/sub-mm morphology of the individual galaxies in detail, based
on the \textit{Herschel} maps and major axis profiles. In
Section~{\ref{VerticalProfiles.sec}} we discuss the vertical
distribution of the FIR/sub-mm emission and look for evidence of cool
interstellar dust at high galactic latitudes. In
Section~{\ref{DustMasses.sec}} we apply a simple modified black-body
fitting to the global fluxes to derive dust masses and temperatures,
and we compare these dust masses with those obtained from radiative
transfer fits to optical data. Finally, in
Section~{\ref{Conclusions.sec}} we present our conclusions.

\section{Observations and data reduction}
\label{Observations.sec}

\subsection{Sample selection}
\label{SampleSelection.subsec}

\begin{table*}[t] 
  \centering
  \caption{Properties of the galaxies in our sample. The distances $D$ are taken from NED, details
    on the specific adopted distance for each galaxy can be found in
    \ref{SampleSelection.subsec}. The second to last column gives the
      conversion between angular and linear scales at the assumed
      distances. The last column gives the inclination of the galaxy,
      derived from radiative transfer modelling (see text for details).} 
  \label{Sample.tab}
  \begin{tabular}{ccccccccc}
    \hline \hline
    galaxy & RA & dec & type & $M_V$ & PA & $D$ & scale
    & $i$ \\
    & (J2000) & (J2000) & & (mag) & (deg) & (Mpc) & (pc/arcsec) & (deg) \\
    \hline
    NGC\,973 & 02:34:20 & +32:30:20 & Sbc & 13.6 & 48.4 & 63.5 & 308 & 89.6 \\
    UGC\,4277 & 08:13:57 & +52:38:54 & Sc & 14.9 & 109.5 & 76.5 & 371 & 88.9 \\
    IC\,2531     & 09:59:56 & --29:37:04 & Sc & 12.9 & 75.7 & 36.8 & 178 & 89.6 \\
    NGC\,4013 & 11:58:31 & +43:56:48 & Sb & 12.1 & 244.8& 18.6 & 90 & 89.7 \\
    NGC\,4217 & 12:15:51 & +47:05:30 & Sb & 12.0 & 49.3 & 19.6 & 95 & 88.0 \\
    NGC\,5529 & 14:15:34 & +36:13:36 & Sc & 12.9 & 294.1 & 49.5 & 240 & 86.9 \\
    NGC\,5907 & 15:15:54 & +56:19:44 & Sc & 11.1 & 154.7 & 16.3 & 79 & 87.2 \\
    \hline
  \end{tabular}
\end{table*}

The \textit{HER}OES sample consists of seven galaxies, which were
  selected from a large sample of edge-on spiral galaxies according to
  the following criteria.  The first criterion is an optical diameter
of at least 4~arcmin, in order to have sufficient spatial resolution
in the \textit{Herschel} images, even at 500~$\mu$m (where the
angular resolution is 36\arcsec). The second criterion is somewhat
more subjective: we require the galaxies to have a clear and regular
dust lane. This requirement was driven by our ambition to construct
detailed radiative transfer models of the observed galaxies, in order
to compare the predicted FIR emission with the observations. This
second requirement limits the sample to galaxies with an inclination
that deviates at most a few degrees from exactly
edge-on. Unfortunately, in the range so close to edge-on, galaxy
inclinations cannot be easily determined in an objective way, for
example based on axial ratios. This criterion also limits the sample
to rather massive galaxies with rotational velocities
$v_{\text{rot}}\gtrsim120$ km\,s$^{-1}$, as less massive galaxies tend
not to show a dust lane even if their inclinations are almost
perfectly edge-on \citep{2004ApJ...608..189D}.\footnote{A
  complementary programme to \textit{HER}OES, the New HErschel
  Multi-wavelength Extragalactic Survey of Edge-on Spirals (NHEMESES)
  is devoted to the study of the interstellar dust medium in a set of
  galaxies with $v_{\text{rot}}<120$ km\,s$^{-1}$
  \citep{2012IAUS..284..128H, 2012A&A...541L...5H}.}

Based on these considerations, the starting point of our sample was
the combination of the samples from \citet{1997A&A...325..135X,
  1999A&A...344..868X} and \citet{2007A&A...471..765B}, since they
were successful in fitting radiative transfer models to optical and
NIR data with their respective codes. A number of the galaxies in this
sample (NGC\,4302, NGC\,5746 and NGC\,5965) showed evidence of
irregular dust lanes and rendered modelling of the extinction very
difficult to impossible, therefore they were omitted from the
sample. The cut in optical diameter removed IC\,1711 from the
sample. Finally, the prototypical example of an edge-on galaxy,
NGC\,891, was omitted as it was slated to be observed by
\textit{Herschel} as part of the Very Nearby Galaxy Survey
(VNGS) key programme.

The remaining seven edge-on spiral galaxies form the sample for the
\textit{HER}OES project (see Table~\ref{Sample.tab} for the
main properties). Distances were taken from NED according to the
following criterion: if the galaxy has a redshift $z > 0.01$ (which is
the case for NGC\,973 and UGC\,4277), we assume the Hubble flow
contribution is dominant and adopt the redshift dependent distance
based on the velocity with respect to the 3K CMB, using $H_{0} =
73$~km~s$^{-1}$~Mpc$^{-1}$; for all others we use the average value of
the redshift independent distance measurements, which are mostly based
on the Tully-Fisher relation. The position angles listed in
Table~{\ref{Sample.tab}} were either taken from
\citet{2007A&A...471..765B}, or determined from the PACS 100~$\mu$m\
images.

\subsection{Herschel observations}

All galaxies were observed with PACS and SPIRE separately, both with
their nominal scan speed, i.e. 20\arcsec/s for PACS and 30\arcsec/s
for SPIRE. Map sizes vary between $8\arcmin~\times~8\arcmin$ and
$16\arcmin~\times~16\arcmin$ and have been chosen to cover enough
surrounding sky area to characterise the background. For the PACS
maps, four cross-scans (i.e. four nominal and four orthogonal scans)
were performed, while the SPIRE maps were observed with a single
cross-scan. 

Full details on all \textit{Herschel} observations carried out can
be found in Table~{\ref{Observations.tab}}.

\begin{table*}
  \centering
  \caption{Details of the \textit{Herschel} observations. OD indicates the
    operational day of the \textit{Herschel} Space Observatory, ObsId is the
    observation identification number. The PACS observations always 
    consist of two concatenated observing blocks, one for the nominal 
    and the other for the orthogonal scan directions. The indicated duration of the
    observations includes both on-source integration time and
    instrument and observation overheads.}
  \label{Observations.tab}
  \begin{tabular}{ccccccc}
    \hline \hline
    OD	&	target		&	instrument	&	duration	&	start date and time &	ObsId	&	field size \\
    & & & (s) & (UT) & & (arcmin$^2$) \\
    \hline
    395	&	NGC\,4013      &	SPIRE &	349	&	2010-06-12~16:29:39	&	1342198241	&	$8\times8$ \\
    467	&	NGC\,5907      &	SPIRE &	770	&	2010-08-24~00:47:30	&	1342203599	&	$16\times16$ \\  
    500	&	UGC\,4277      &	SPIRE &	349	&	2010-09-26~16:16:14	&	1342205086	&	$8\times8$ \\ 
    558	&	NGC\,4217      &	SPIRE &	541	&	2010-11-22~21:56:09	&	1342210500	&	$11\times11$ \\
    558	&	IC\,2531          &	SPIRE  &	529	&	2010-11-23~05:49:48	&	1342210524	&	$10\times10$ \\
    572	&	NGC\,5529      &	SPIRE &	529	&	2010-12-07~11:06:03	&	1342210882	&	$10\times10$ \\ 
    715	&	NGC\,5907      &	PACS  &	2124&	2011-04-29~13:57:35	&	1342220804	&	$16\times16$ \\  
    715	&	NGC\,5907      &	PACS  &	2124&	2011-04-29~14:34:02	&	1342220805	&	$16\times16$ \\  
    723	&	NGC\,4217      &	PACS  &	1217&	2011-05-07~11:07:33	&	1342220109	&	$11\times11$ \\
    723	&	NGC\,4217      &	PACS  &	1217&	2011-05-07~11:28:53	&	1342220110	&	$11\times11$ \\
    723	&	UGC\,4277      &	PACS  &	813	&	2011-05-07~12:43:07	&	1342220119	&	$8\times8$ \\ 
    723	&	UGC\,4277      &	PACS  &	813	&	2011-05-07~12:57:43	&	1342220120	&	$8\times8$ \\ 
    731	&	NGC\,4013      &	PACS  &	813	&	2011-05-15~12:13:15	&	1342220968	&	$8\times8$ \\
    731	&	NGC\,4013      &	PACS  &	813	&	2011-05-15~12:27:51	&	1342220969	&	$8\times8$ \\ 
    733	&	IC\,2531	        &	PACS  &	909	&	2011-05-17~05:29:37	&	1342221271	&	$10\times10$ \\
    733	&	IC\,2531	        &	PACS  &	909	&	2011-05-17~05:45:49	&	1342221272	&	$10\times10$ \\ 
    758	&	NGC\,5529      &	PACS  &	909	&	2011-06-11~17:40:01	&	1342222509	&	$10\times10$ \\ 
    758	&	NGC\,5529      &	PACS  &	909	&	2011-06-11~17:56:13	&	1342222510	&	$10\times10$ \\ 
    787	&	NGC\,973	&	PACS  &	813	&	2011-07-10~10:55:12	&	1342223868	&	$8\times8$ \\ 
    787	&	NGC\,973	&	PACS  &	813	&	2011-07-10~11:09:48	&	1342223869	&	$8\times8$ \\ 
    828	&	NGC\,973         &	SPIRE &	349	&	2011-08-20~10:49:43	&	1342226629	&	$8\times8$ \\ 
    \hline
  \end{tabular}
\end{table*}

\subsection{Data reduction}
\label{DataReduction.subsec}

For the PACS data reduction, the \textit{Herschel} Interactive
Processing Environment \citep[HIPE,][]{2010ASPC..434..139O} v8.0 with
PACS Calibration version 32 was used to bring the raw data to Level-1,
which means flagging of pixels, flat-field correction, conversion to
Jansky and assigning sky coordinates to each detector array
pixel. These intermediate timelines were then fed into
\texttt{Scanamorphos} v15, an IDL program which is capable of removing
the 1/f noise, drifts and glitches by using the redundancy in the
observational data and in the end produces the resulting maps \citep{2012arXiv1205.2576R}. The pixel sizes in the final maps are
the default values used in \texttt{Scanamorphos}, i.e. 1\farcs70 and
2\farcs85 for the 100~$\mu$m\ and 160~$\mu$m\ maps respectively. These
pixel sizes correspond to one quarter of the point spread function
(PSF) full width at half maximum (FWHM) for the scan speed used in our
observations, which has values of 6\farcs8 and 11\farcs4 at 100
  and 160~$\mu$m respectively \citep{PACSOM}. In producing the final
maps, a customised sky grid was used to project the major axes
horizontally.

To process the SPIRE data up to Level-1, which was done using HIPE
v8.0, the official pipeline was modified into a custom script,
applying the latest calibration products (calibration tree
v8.1). Instead of applying the default temperature drift correction
and median baseline subtraction, we used a custom method called {\sc
  BriGAdE} (Smith et al. in prep.) to correct for the temperature
drifts. The final SPIRE maps were produced with the naive mapper,
using pixel sizes of 6\arcsec, 8\arcsec and 12\arcsec for the
250~$\mu$m, 350~$\mu$m\ and 500~$\mu$m\ maps respectively.  These
sizes were chosen to measure about a third of the SPIRE beams' FWHM,
having values of 18\farcs2, 24\farcs5 and 36\farcs0 respectively
\citep{SPIREOM}. As was the case for the PACS maps, the final SPIRE
maps were projected onto a sky grid with a horizontal major axis.

Due to some internal calibration adjustments on the satellite,
  observations performed during operational day (OD) between 320 and
  761 might suffer from pointing accuracy issues, and may not always
  be trustable. All of our observations, apart from those for
  NGC\,973, could then be affected by potential astrometry issues. We
  have taken this into account in the data reduction by using the
  pointing product supplied by the
  HSC\footnote{\href{http://herschel.esac.esa.int/twiki/bin/view/Public/HowToUseImprovedPointingProducts}{\texttt{http://herschel.esac.esa.int/twiki/bin/view/Public/\\*HowToUseImprovedPointingProducts}}}.

\subsection{Optical Images}

V-band images used for this work are a combination of both data
already available in the literature and new observations. UGC\,4277,
NGC\,5529, NGC\,4013 and NGC\,4217 are from \citet{2008A&A...490..461B} 
(see section 2 of that paper for a detailed presentation of these
data). Data for IC\,2531 are taken from \citet{1999A&A...344..868X} (a
detailed description is given in their section 2). Finally, data for
NGC\,5907 and NGC\,973 are from newly performed observations at the
Telescopio Nazionale Galileo (TNG). Both galaxies were observed in 5
exposures of 150 seconds between August and October 2011, with a
seeing of about $1.5''$. As NGC\,973 is located close to a bright
star, causing severe reflection effects on the first observations, it
was re-observed with a rotated field of view to avoid the bright
star. NGC\,5907 is larger than the TNG-DOLORES field of view, so that
two different observations had to be combined.

Optical images are shown here (see Figures~{\ref{NGC973.fig}} to
{\ref{NGC5907.fig}}) mainly for illustrative purposes, as they will be
exploited in future works in the frame of the \textit{HER}OES
project.

\section{Global fluxes}
\label{GlobalFluxes.sec}

\begin{table*}
  \caption{Global PACS and SPIRE fluxes and the major and minor 
    semi-axes of the used elliptical apertures for the galaxies in the
    sample.  All fluxes include a colour correction.}
\label{GlobalFluxes.tab}
\centering
\begin{tabular}{lcccccrr}
\hline \hline
galaxy & $F_{100}$ & $F_{160}$ & $F_{250}$ & $F_{350}$ & $F_{500}$ & a & b \\
& [Jy] & [Jy] & [Jy] & [Jy] & [Jy] & $['']$  & $['']$  \\
\hline
NGC\,973   & $     4.04\pm0.20$  & $    4.92\pm0.24$ &  $   3.84\pm 0.41$  & $    1.81\pm 0.28$   & $   0.69\pm0.17$   & 144 &   71  \\
UGC\,4277 & $    1.46\pm0.09$   & $   2.89\pm0.13$ & $    2.48\pm 0.34$  & $   1.38\pm 0.24$   &  $   0.58\pm0.18$   & 139 &   74  \\
IC\,2531      & $    5.22\pm0.26$   &$ 10.07\pm0.50$ & $    6.70\pm 0.64$  & $   3.62\pm 0.40$   &  $  1.57\pm0.24$   & 240 &   78  \\
NGC\,4013 & $ 22.36\pm1.12$   & $ 32.04\pm1.60$ & $ 18.27\pm 1.34$  & $    7.93\pm 0.63$   & $   2.77\pm0.32$   & 192 &    95 \\
NGC\,4217 & $ 38.69\pm1.94$   & $ 56.79\pm2.84$ & $ 28.79\pm 2.07$  & $ 12.27\pm 0.94$   & $    4.25\pm0.39$   & 216 & 118 \\
NGC\,5529 & $   7.91\pm0.40$   & $ 12.89\pm0.65$ & $    8.42\pm 0.73$  & $   4.25\pm 0.47$   & $    1.74\pm0.27$   & 241 &   87  \\
NGC\,5907 & $ 57.35\pm2.87$   & $ 88.56\pm4.43$ & $ 54.21\pm 3.93$  & $ 25.83\pm 1.98$   & $ 10.11\pm0.87$   & 386 & 112 \\
\hline
\end{tabular}
\end{table*}
 
We determined the global flux of the seven \textit{HER}OES
galaxies in the two PACS and three SPIRE bands using aperture
photometry. We adopted the same approach as described in
\citet{2012ApJ...745...95D} for the flux determination, background
subtraction, and uncertainties calculation. To derive the total fluxes
we defined, for each galaxy, an elliptical aperture, roughly centred
on the galaxy itself. For each galaxy, the size of the aperture is
defined in such a way that it encompasses all the emission at 500
$\mu$m, as this band has the poorest spatial resolution.  No
  aperture corrections were applied to the measured fluxes.  In any case,
  using the image at the poorest resolution to define the ellipses
  within which fluxes are calculated should ensure that no flux is
  lost. As a further check, we performed flux measurements on the
  500 $\mu$m images by choosing various apertures at increasingly
  higher $a$ and $b$ (major and minor axis of the ellipses,
  respectively) values. The results make us confident that even at the
  longest wavelength we are not missing any significant fraction of
  the extended emission.

The lengths of the major and minor semi--axes are reported in
Table~{\ref{GlobalFluxes.tab}} for each galaxy. We then used the
DS9/Funtools program {\sc{Functs}}, which sums the flux value of each
pixel inside the aperture, to yield the global flux.  Sky subtraction
is performed as follows: we define a number of circular apertures
surrounding the galaxy, far enough so that they are not contaminated
by its emission, but as close as possible to the galaxy in order to
better sample the sky properties of its surroundings, also making sure
they do not include any background source. A mean sky level per pixel
is then computed from these apertures and this value, multiplied by
the number of pixels encompassing the galaxy within the elliptical
aperture, is subtracted from the global flux measured as explained
before.

A colour correction factor is included, taken from the ICC ``PACS
  Photometer - Colour Corrections'' document\footnote{PACS Photometer
    Passbands and Colour Correction Factors for Various Source SEDs,
    April 12, 2011} and from the ``SPIRE observers Manual''
  \footnote{Version 2.4, June 7, 2011}, respectively. For PACS, we
  used the tabulated values relative to a modified black body
  with temperature of 20 K and emissivity index $\beta=2$, which are
  both well suited for the SEDs of our objects (see
  Section~\ref{ssect:sed}; values relative to T=15 K were adopted for
  UGC\,4277 as its best fit model favours a lower temperature).  As for
  SPIRE, we used colour correction values calculated for extended
  sources, with an index $\alpha_S=4$ (which turned out to be the most
  appropriate choice of $\beta=1.8$). A calibration correction factor
  was also applied to account for the fact that our sources are
  extended (see SPIRE Observer's Manual, Section~5.2.8).

As for the uncertainties, we calculate the total uncertainty by
summing in quadrature the calibrations uncertainty (5 and 7 \% for
PACS and for SPIRE, respectively\footnote{See ``PACS Observer's
  Manual'', Version 2.3, and ``SPIRE Observer's Manual'', Version
  2.4}) and the uncertainty on the sky level, which we compute as in
Eq. (2) in \citet{2012ApJ...745...95D}.

As a check on the accuracy of the derived flux densities (fluxes are
reported in Table~{\ref{GlobalFluxes.tab}}), we searched in the
literature and various archives for other flux densities at similar
wavelengths. For all galaxies but IC\,2531, \textit{IRAS} flux
densities are available at 100~$\mu$m\ \citep{1990BAAS...22Q1325M,
  2003AJ....126.1607S, 2007A&A...462..507L}. All galaxies except
NGC\,973 and UGC\,4277 were detected by \textit{Planck} in the
Early Release Compact Source Catalog \citep{2011A&A...536A...7P}. The
two brightest galaxies in our sample, NGC\,5907 and NGC\,4217, were
detected by \textit{ISO} as part of the ISOPHOT 170~$\mu$m\
Serendipity Survey \citep{2004A&A...422...39S}. Finally, four galaxies
(NGC\,973, NGC\,4217, NGC\,5529 and NGC\,5907) are listed in the
\textit{Akari}/FIS All-Sky Survey Point Source Catalogue
\citep{2010yCat.2298....0Y}.

\begin{figure} 
  \centering
  \includegraphics[width=\columnwidth]{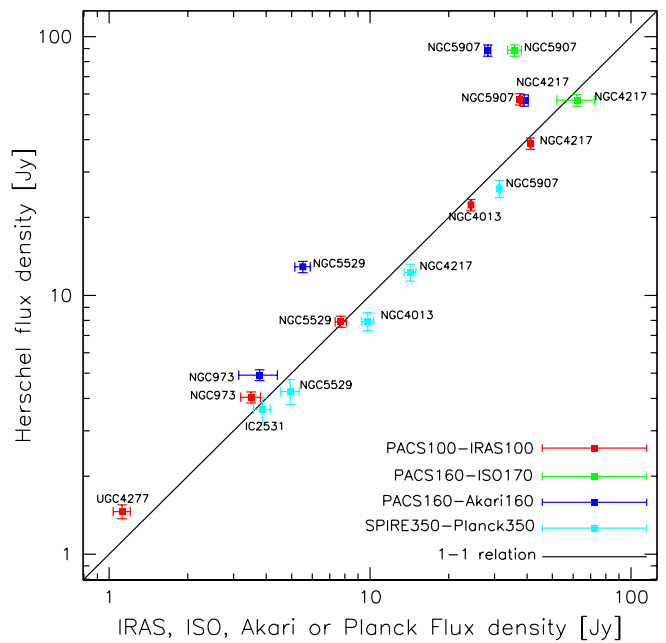}
  \caption{Comparison between the PACS 100~$\mu$m, PACS 160~$\mu$m\
    and SPIRE 350~$\mu$m\ fluxes with \textit{IRAS},
    \textit{ISO}, \textit{Akari} and
    \textit{Planck} fluxes.}
\label{FluxComparison.fig}
\end{figure} 

Figure~{\ref{FluxComparison.fig}} shows the comparison of the PACS
100~$\mu$m, PACS~160~$\mu$m\ and SPIRE 350~$\mu$m\ fluxes of the
\textit{HER}OES galaxies with the \textit{IRAS}
100~$\mu$m, \textit{Akari} 160~$\mu$m, \textit{ISO}
170~$\mu$m\ and \textit{Planck} 857~GHz fluxes
respectively. The solid line in the plot shows the one-to-one
relationship. The agreement between the \textit{Herschel}
fluxes and the archival \textit{IRAS}, \textit{ISO} and
\textit{Planck} (aperture) fluxes is generally excellent. The
only exception is NGC\,5907, where the PACS 100 and 160~$\mu$m\ fluxes
deviate significantly from the \textit{IRAS} 100~$\mu$m\ and
ISOPHOT 170~$\mu$m\ fluxes respectively: 57.35 versus 37.43~Jy at
100~$\mu$m\ and 88.56 versus 35.83~Jy at 160/170~$\mu$m. A possible
explanation is the large extent of this galaxy, which is resolved even
by the 2.94\arcmin\ \textit{IRAS} 100~$\mu$m\ beam
\citep{1989AJ.....98..766S}.  The SPIRE 350~$\mu$m\ flux
($25.83\pm1.98$~Jy) and \textit{Planck} 857~GHz flux
($27.25\pm0.69$~Jy) of NGC\,5907 are in excellent agreement.

While the agreement between \textit{IRAS}, \textit{ISO}
and \textit{Planck} on the one hand, and
\textit{Herschel} on the other hand is very good,
Figure~{\ref{FluxComparison.fig}} shows that the agreement between the
PACS 160~$\mu$m\ flux densities and the \textit{Akari}/FIS
160~$\mu$m\ flux densities is very poor. The PACS fluxes are on
average a factor two higher than the \textit{Akari}
fluxes. This difference is probably due to the way
\textit{Akari} fluxes are measured, i.e. similarly to PSF
fitting. While the PSF of \textit{Akari} is about $60''$ at
160~$\mu$m, all the galaxies in our sample are much larger, so it is
indeed expected that the latter are smaller than the
\textit{Herschel} fluxes.

\section{FIR/sub-mm morphology}
\label{Morphology.sec}

In this section we describe the far-infrared morphology of the
observed galaxies, based on the PACS and SPIRE maps presented in
Figures~{\ref{NGC973.fig}} through {\ref{NGC5907.fig}}. We also
produce horizontal profiles of the FIR/sub-mm emission in each of the
\textit{Herschel} bands, which we present in
Figure~\ref{HorizontalProfiles.fig}. These profiles were derived by
integrating the flux over the image pixels along strips parallel to
the minor axis and averaging through division by the number of pixels
used for each strip. The area used for this procedure was a rectangle
centred on the galaxy, having sizes equal to the major and minor axes
of the elliptical apertures used in the flux determination
(Table~\ref{GlobalFluxes.tab}).  These profiles are hence a sort of
average surface brightness profiles along the horizontal direction. We
used these collapsed profiles instead of straight cuts along the major
axis to increase the signal-to-noise. 

We compare the observed profiles to the theoretical horizontal
  one expected from the so-called double-exponential model,
  described by the three-dimensional density distribution
  \begin{equation} 
    \rho(R,z) 
    =
    \frac{M_{\text{d}}}{4\pi\,h_R^2\,h_z}
    \exp\left(-\frac{R}{h_R}-\frac{|z|}{h_z}\right) 
    \label{ded} 
  \end{equation}
  with $R$ and $z$ the horizontal and vertical position within the
  system's coordinates, $M_{\text{d}}$ the total dust mass, and $h_R$
  and $h_z$ the dust's horizontal scalelength and vertical
  scaleheight, respectively. This model is the most commonly adopted
  description of the three-dimensional distribution of dust in spiral
  galaxies \citep[e.g.][]{1997A&A...325..135X, 1998A&A...331..894X,
    1999A&A...344..868X, 2004A&A...425..109A, 2007A&A...471..765B,
    2008A&A...490..461B, 2011A&A...527A.109P}.  If this system is
  perfectly edge-on, with the major axis aligned to the $x$-axis, it
  has a mass surface density distribution which is described by the 
  following:
\begin{equation}
  \Sigma(x,y)
  =
  \frac{M_{\text{d}}}{2\pi\,h_R\,h_z}
  \left(\frac{|x|}{h_R}\right)\,
  K_1\left(\frac{|x|}{h_R}\right)
  \exp\left(-\frac{|y|}{h_z}\right)
\label{ded-Sigma}
\end{equation}
where $K_1$ is the modified Bessel function of the first order
\citep[see e.g.][Equation~2]{2002MNRAS.334..646K}.  When we collapse this
surface density distribution in the vertical direction and normalise
the resulting expression, we find as horizontal profile:
\begin{equation}
  \Sigma_{\text{hor}}(x) 
  = 
  \frac{1}{\pi\,h_R} 
  \left(\frac{|x|}{h_R}\right) 
  K_1\left(\frac{|x|}{h_R}\right)
  \label{dust-mass-radial-profile}
\end{equation}

In Figure~\ref{HorizontalProfiles.fig} we also show these exponential
models with scalelengths determined from modelling of optical data
taken from the literature (see Table~\ref{DustMasses.tab}),
represented with black crosses.  However, it should be pointed out
that these models only describe the dust distribution, while the
actual data profiles from the \textit{Herschel} maps show instead the
characteristics of the dust emission, such
  as its temperature.  For most galaxies in our sample, the
500~$\mu$m profiles follow this exponential model to a reasonable
degree, although even this model is incomplete: it fails to include
the truncation of the dust disc.  Vertical profiles were derived in a
similar fashion; these are presented in
Figure~\ref{VerticalProfiles.fig} and discussed in Section
\ref{VerticalProfiles.sec}.

Information such as distances, optical radius, morphological and
spectral types are taken from the NASA Extragalactic Database (NED).

\subsection{NGC\,973 [UGC\,2048]}
\label{NGC973.sec}

\begin{figure*}
  \centering
  \includegraphics[width=\textwidth]{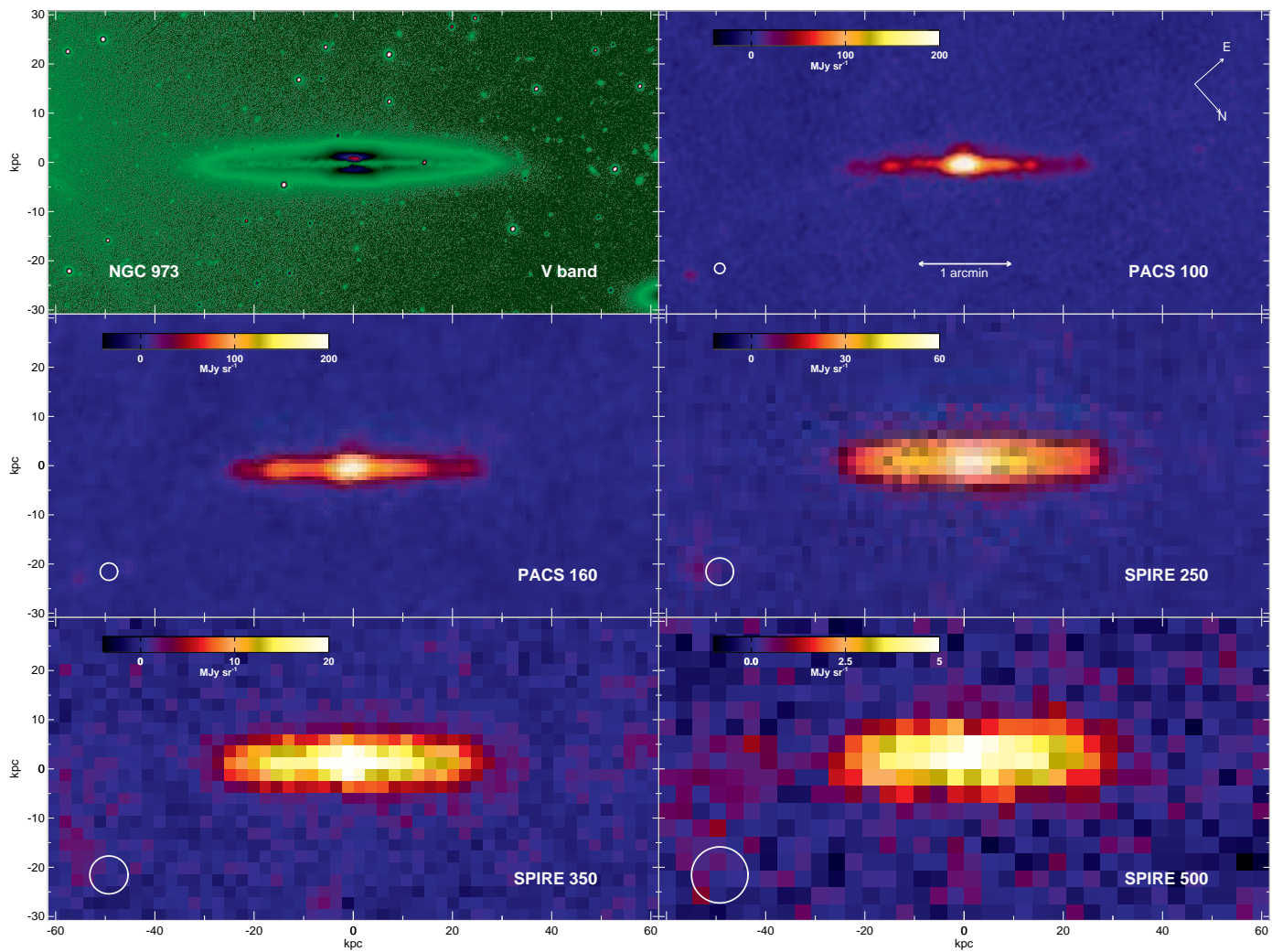}
  \caption{Optical V-band and \textit{Herschel} PACS and SPIRE images for
    NGC\,973. The image sizes are about $6' \times 3'$.  The orientation and length scale are given in the top right panel.  The band for each panel is shown in its bottom right corner, with the \textit{Herschel} beam size indicated by the circle in its bottom left corner.}
  \label{NGC973.fig}
\end{figure*}

NGC\,973 is located at a distance of 63.5~Mpc and has a $D_{25}$
diameter of 4\farcm03. Its morphological class is estimated to be Sb
or Sbc, due to the inherent uncertainty of the precise structure of
edge-on galaxies, and it is classified as a Seyfert~2/LINER
galaxy. The inclination is found to be $89\fdg6\pm0\fdg1$
\citep{1997A&A...325..135X} and the galaxy is showing a prominent dust
lane \citep{1992A&AS...93..255G} over its entire span, from the centre
to the edges. Although \citet{1994ApJ...427..160P} claimed not to have
found any extraplanar emission-line gas in H$\alpha$,
\citet{2003ApJS..148..383M} did detect some extraplanar emission from
gas, but also mentioned there is no suggestion of a widespread diffuse
emission above the disc plane.  It is assumed the dust shows the same
or at least similar behaviour. A previous attempt to model the dust
distribution was presented in \citet{1997A&A...325..135X}, resulting
in a dust scalelength of 50\arcsec or 16.3~kpc, about 50\% longer than
the equivalent stellar disc scalelength.

The different wavelength maps, horizontal and vertical profiles for
NGC\,973 are shown in Figures~\ref{NGC973.fig},
\ref{HorizontalProfiles.fig} and \ref{VerticalProfiles.fig}
respectively.  This galaxy demonstrates a very peculiar behaviour in
its horizontal profiles compared to the other galaxies in the
sample. For one thing, the surface brightness shows a relatively
shallow decline, out to about 1\farcm4 on the SW side and about
1\farcm4 to 1\farcm6 -- depending on the wavelength -- on the NE side
of the disc and then falls down dramatically, suggesting a possible
truncation of the dust disc. In addition to this, the emission shows a
very sharp and strong central peak at PACS wavelengths, which has
almost completely vanished from 350~$\mu$m\ onwards, partly because of
the poorer spatial resolution at these latter wavebands. This strong
peak is overshadowing the other secondary peaks visible at PACS
wavelengths on either side (at roughly 0\farcm75 and --to a lesser
extent-- 1\farcm2 from the centre).

The central peak clearly visible in both the 100 and 160~$\mu$m\
horizontal profiles might be linked to the fact that this galaxy hosts
an active galactic nucleus, which would then be responsible for the
compact, warmer emission in the innermost regions.  We
  checked whether this peak is compatible with a point-like
  emission, as expected --at these resolutions-- for an AGN--like
  source. To do so, we convolved a set of gaussians with the
  1-dimensional PSF profile derived from PACS PSF
  \citep{2011PASP..123.1218A}, and subtracted them from the horizontal
  profile derived at 100~$\mu$m, the highest resolution data. 
  We found that the central peak is compatible with a
  gaussian profile emission with an upper limits on the central
  source's FWHM of $1\farcs5$, corresponding to a physical scale of
  $\sim 450$ pc. Due to the low spatial resolution at SPIRE
wavelengths it is very difficult to draw any further conclusions about
the disc structure.

\subsection{UGC\,4277}

\begin{figure*} \centering
  \includegraphics[width=\textwidth]{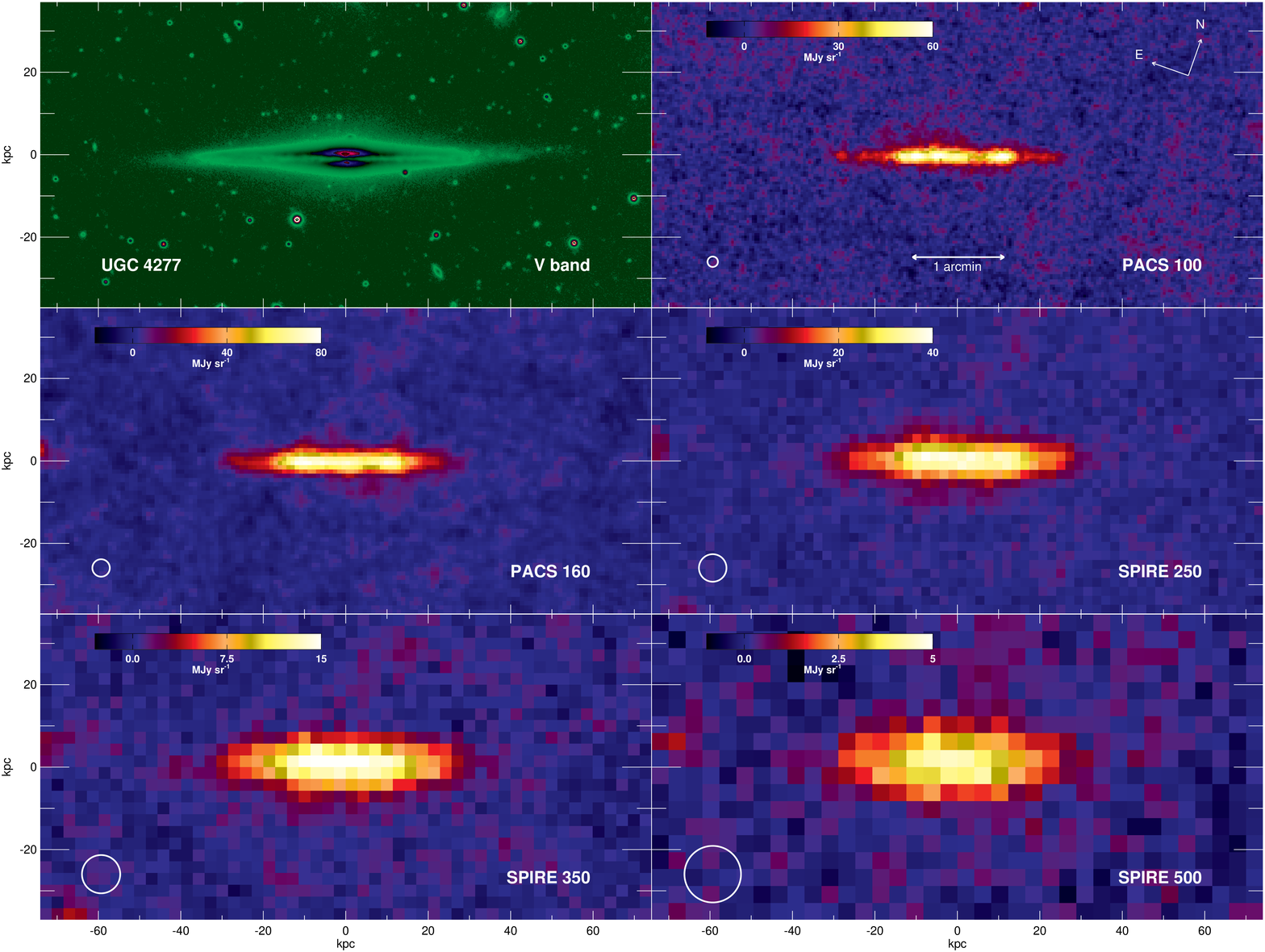}
  \caption{Optical V-band and \textit{Herschel} PACS and SPIRE images for
    UGC\,4277. The image sizes are about $6' \times 3'$.}
\label{UGC4277.fig}
\end{figure*}

UGC\,4277 is the most distant galaxy in our sample at a distance of
76.5~Mpc, with a $D_{25}$ diameter of 3\farcm9 and a classification as
an Sc or Scd type galaxy. It has an inclination of $88\fdg89$
\citep{2007A&A...471..765B}, with a clear dust lane visible from the
centre to the edges. Modelling of the dust distribution has been
previously carried out in \citet{2007A&A...471..765B}, giving a dust
scalelength of 35\arcsec or 12.5~kpc, which is about the same as the
fitted stellar disc scalelength.

The horizontal profiles for UGC\,4277 -- maps are given in
Figure~\ref{UGC4277.fig} -- show a rather erratic behaviour.  At 100
$\mu$m, a number of small peaks are visible, but none of these can be
pinned down as the central one. On top of that, a smaller secondary
peak about 1\farcm3 off centre toward the east end of the galaxy is
almost separated from the rest of the disc by a clear drop in emission
at this wavelength and emission levels fall sharply just beyond this
peak, while on the other side, the profile gets dominated by the
background somewhere around 1\farcm5 away from the centre. The other
PACS wavelength is more symmetrical, with one central peak and one
secondary about 0\farcm5 off centre on either side -- though all three
have roughly the same strength -- and the profile drops off very
steeply beyond these secondaries, but first levels out at a reduced
plateau before falling off farther beyond 1\farcm3 on both sides. At
SPIRE wavelengths, the surface brightness seems to be smoother,
maintaining a plateau up to 0\farcm4 to 0\farcm5 from the centre
before declining down to the background at about 1\farcm6 off centre.

An exponential distribution for the dust turns out to be a fairly good
representation of the observed horizontal profiles at SPIRE
wavelengths while PACS profiles are characterised by a more irregular
behaviour. The only clearly distinguishable feature in the maps is a
concentration of the surface brightness on one side of the centre at
100~$\mu$m, rendering the structure of the dust disc hard to
determine.

\subsection{IC\,2531}

\begin{figure*} \centering
  \includegraphics[width=\textwidth]{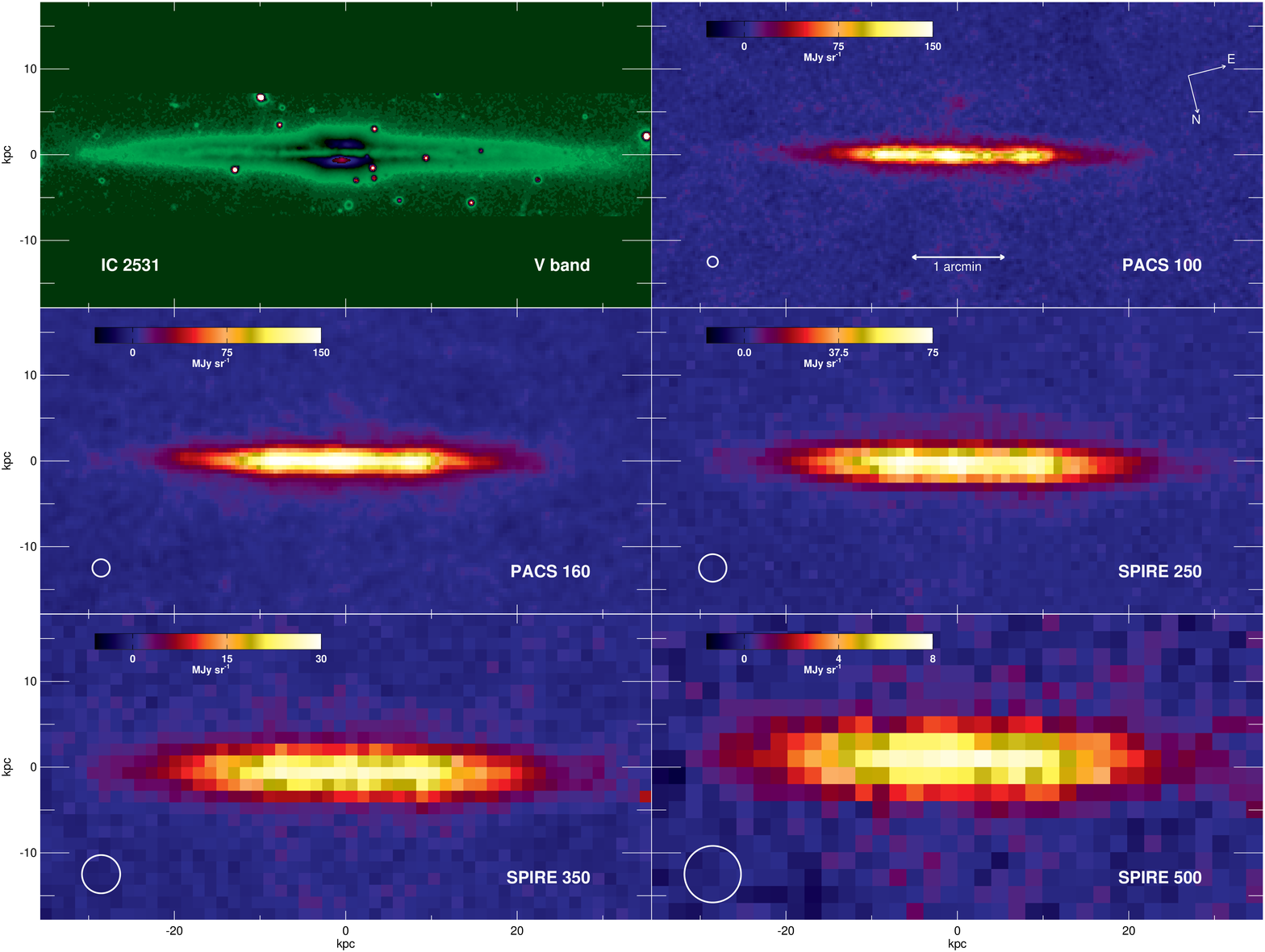}
  \caption{Optical V-band and \textit{Herschel} PACS and SPIRE images for
    IC\,2531. The image sizes are about $6' \times 3'$.}
\label{IC2531.fig}
\end{figure*}

IC\,2531 is the only galaxy on the southern hemisphere in our sample.
It is located at a distance of 36.8~Mpc and it has a $D_{25}$ diameter
of 7\farcm5. It is classified either as an Sb or as an Sc type galaxy
showing H{\sc{ii}} regions, and it is characterised by a conspicuous
peanut-shaped or boxy bulge \citep{1986AJ.....91...65J,
  1987A&AS...70..465D}. With a fitted inclination of
$89\fdg6\pm0\fdg2$ \citep{1999A&A...344..868X}, this galaxy is
oriented almost perfectly edge-on. The dust lane is very prominent and
regular, and extends from the galactic centre to the edge of the disc,
as well as into the higher stellar layers, which is evidenced by dust
features traceable up to a few scaleheights
\citep{2000MNRAS.313..800D}. Many attempts have been made to determine
the dust distribution from fitting extinction models to the dust lane
by several works including \citet{1989ApJ...337..163W},
\citet{1996A&A...309..715J}, \citet{1998AJ....115.1438K} and
\citet{1999A&A...344..868X}, but only the latter performed a global
fit. According to the model by \citet{1999A&A...344..868X}, which
includes both absorption and scattering, the dust in IC\,2531 has a
horizontally extended distribution with a scalelength of 75\arcsec or
13.7~kpc, about 50\% larger than the fitted stellar scalelength.

The maps for IC\,2531 are shown in Figure~\ref{IC2531.fig}.  Apart
from the one at 500~$\mu$m, all horizontal profiles show a plateau
with a primary central peak and clear secondary peaks symmetrically
positioned around the centre at a distance of roughly 1\arcmin, beyond
which point they drop off sharply, out to about 2\arcmin to 2\farcm5
from the centre, where they are no longer distinguishable from the
background. Possibly due to the lower resolution, the surface
brightness at 500~$\mu$m\ remains almost flat from the centre out to
about 1\farcm5 on either side, where it starts to drop off, though not
as sharp as in the other bands.

Because of the clear secondary peaks, especially the 100~$\mu$m\ map
indicates a possible ring-like or spiral structure, along with a more
diffuse dust disc, although this could simply be the continuation of
the same structure seen edge-on. It is also not clear whether this is
a single structure or possibly two or more concentric rings -- as
indicated by the slight revival of the 100~$\mu$m\ emission at about
1\farcm5 on either side of the centre.

\subsection{NGC\,4013}

\begin{figure*} \centering
  \includegraphics[width=\textwidth]{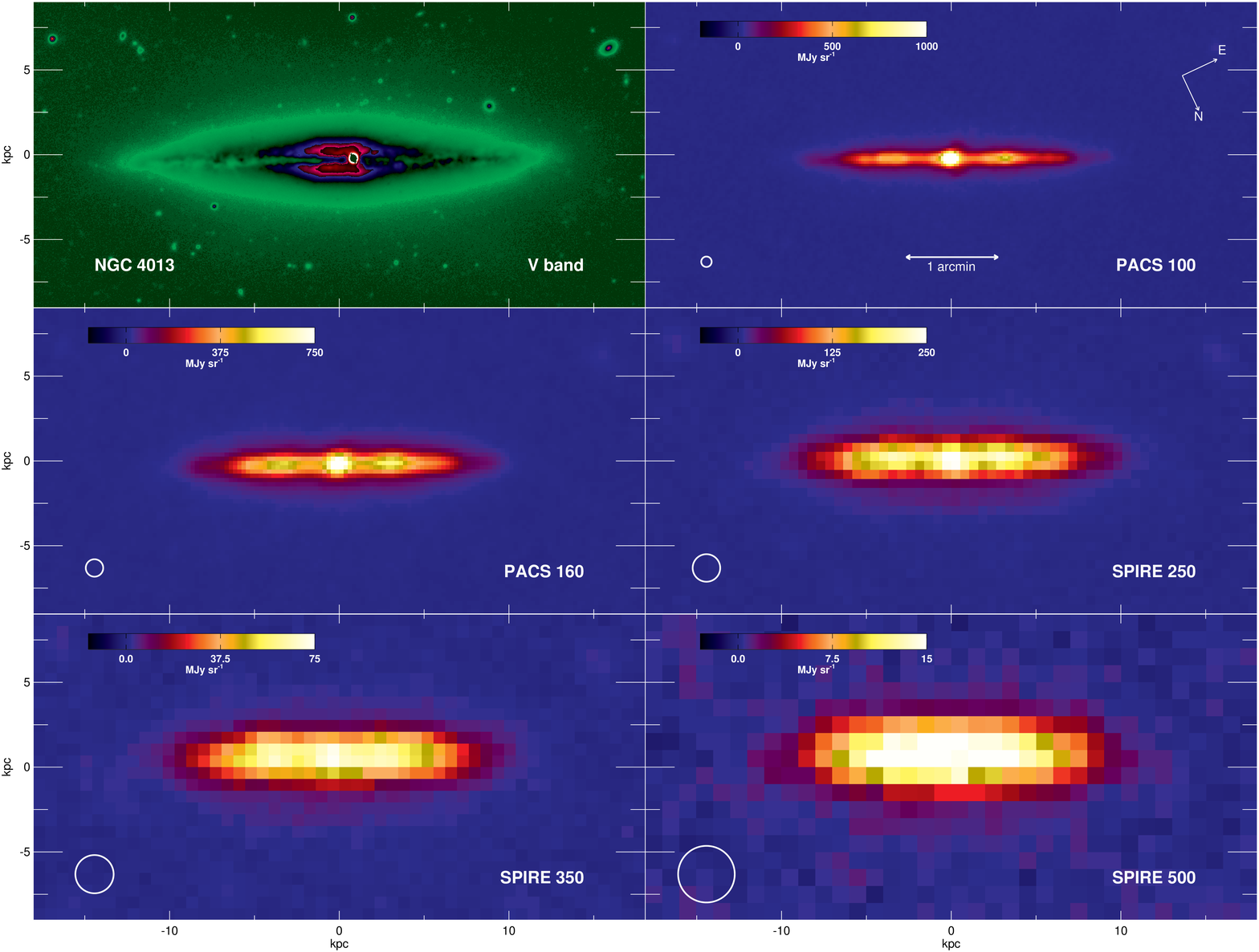}
  \caption{Optical V-band and \textit{Herschel} PACS and SPIRE images for
    NGC\,4013. The image sizes are about $6' \times 3'$.}
\label{NGC4013.fig}
\end{figure*}

At a distance of 18.6~Mpc, NGC\,4013 has a $D_{25}$ diameter of
5\farcm2, is variously classified as an Sbc
\citep{1999A&A...344..868X}, an Sb or an SAb type galaxy with
H{\sc{ii}} regions and is considered a LINER galaxy. Furthermore, the
galaxy possesses a box-shaped bulge \citep{1986AJ.....91...65J,
  1987A&AS...70..465D} and shows a very clear warp in various bands
but sometimes in opposite directions for the dust, gas and stellar
component of the galaxy disc \citep{1987Natur.328..401B,
  1990MNRAS.246..458S, 1991A&A...242..301F, 1995A&A...295..605B,
  1996A&A...306..345B}, although \citet{1999A&A...344..868X} claim the
stellar distribution in their optical images has no warp significant
enough to affect their model. The inclination angle estimates for this
galaxy vary from $89\fdg7\pm0\fdg1$ \citep{1999A&A...344..868X} to
$89\fdg89$ \citep{2007A&A...471..765B}. Apart from a bright foreground
star blocking the view, the dust lane in this galaxy can be clearly
traced from the centre out to the furthest edges as well as out of the
central plane in optical images from \citet{1999AJ....117.2077H},
especially at larger radii where the bulge no longer dominates. Its
dust distribution has been previously modelled with global fits to
optical extinction data in \citet{1999A&A...344..868X} and
\citet{2007A&A...471..765B}, resulting in dust scalelengths varying
from 45\arcsec or 3.9~kpc \citep{1999A&A...344..868X} to 30\arcsec or
2.7~kpc \citep{2007A&A...471..765B}, corresponding to about 25\%
longer \citep{1999A&A...344..868X}, respectively shorter
\citep{2007A&A...471..765B}, than the stellar disc
scalelength. Furthermore, \citet{2011ApJ...738L..17C} have found
evidence of a substantial secondary thick disc and a third stellar
vertically extended disc structure.

The different wavelength maps for NGC\,4013 are presented in
Figure~\ref{NGC4013.fig}.  The horizontal profiles show a very sharp
central peak at 100~$\mu$m, as well as symmetrically located secondary
peaks about 0\farcm6 off centre, although the latter have a difference
in strength, with the one at the NE end of the disc being slightly
brighter. Both primary and secondary peaks stand out above a plateau
level, which reaches out to about 0\farcm6 to 1\arcmin either side of
the centre, while the peaks seem to have evened out at 350~$\mu$m\ and
beyond, most probably due to the coarser resolution. Beyond the
plateau, the profiles become much steeper, up to about 1\farcm3 to
2\arcmin from the centre, where the background begins to dominate.

The horizontal profile displays a substantial departure from the
simple exponential model we superimposed to the data, at PACS
wavelengths, due to the highly peaked emission, while it is more
consistent at the longest wavelengths. The PACS maps are, in fact,
largely dominated by the emission coming from the central region,
possibly contaminated by some nuclear activity as its LINER
classification hints at.  Similarly as for NGC\,973 (see
  Section~{\ref{NGC973.sec}}), we determined an upper limit on
  the extent of the emitting region by convolving a set of gaussians
  with the PACS PSF and subtracting them from the horizontal
  profile. For NGC\,4013, we find an upper limit on the central
  source's FWHM of $1\farcs3$, corresponding to a physical scale of
  $\sim 120$ pc.

Apart from this prominent central peak, the emission from the disc
seems to be concentrated in a single ring-like or spiral
structure. This is much more difficult to discern in the SPIRE maps,
due to the poorer resolution.

\subsection{NGC\,4217}

\begin{figure*} \centering
  \includegraphics[width=\textwidth]{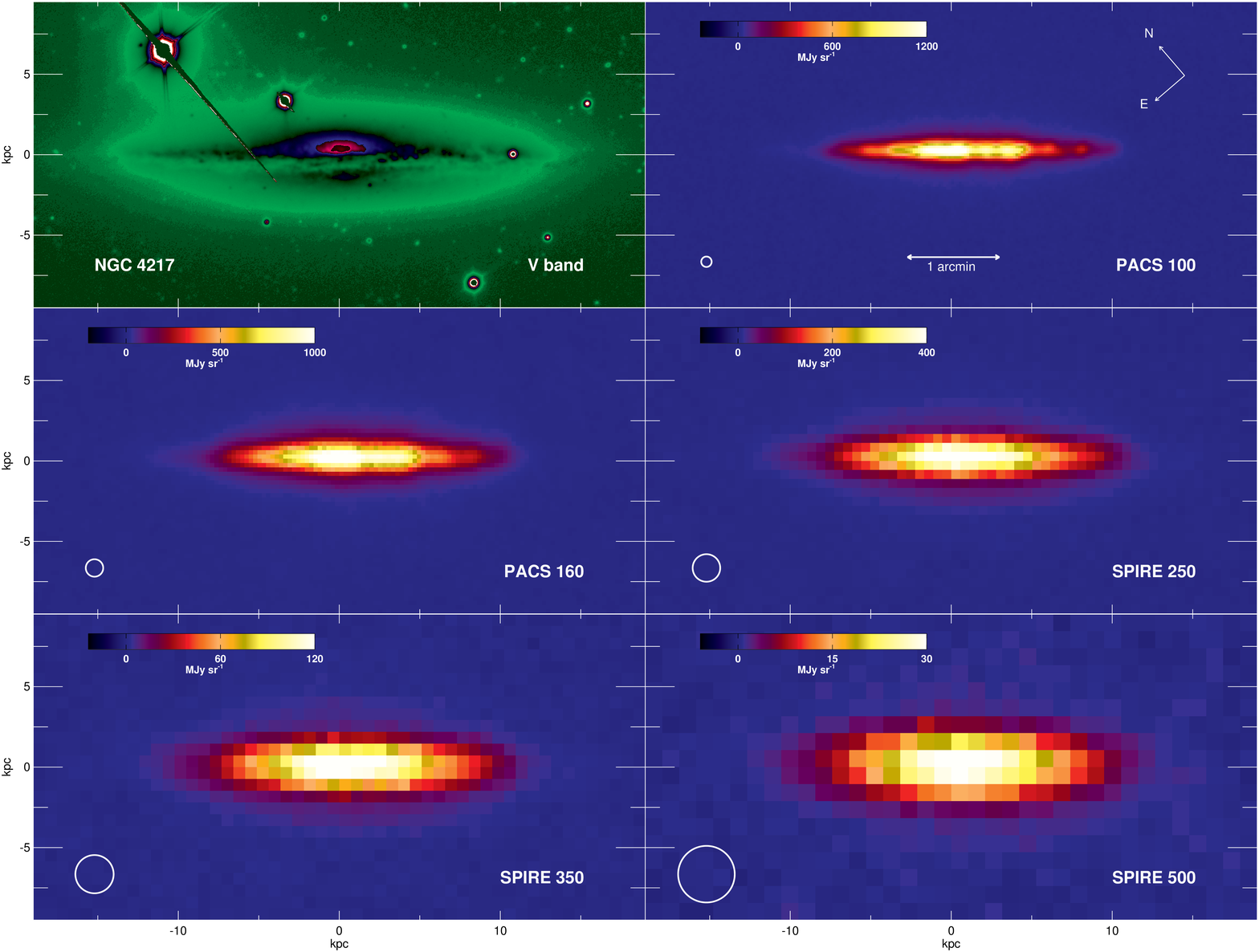}
  \caption{Optical V-band and \textit{Herschel} PACS and SPIRE images for
    NGC\,4217. The image sizes are about $6' \times 3'$.}
\label{NGC4217.fig}
\end{figure*}

For NGC\,4217, we assume a distance of 19.6~Mpc. The galaxy has a
$D_{25}$ diameter of 5\farcm2 and is classified as an Sb or SAb sp
type galaxy, showing H{\sc{ii}} emission regions. Having an
inclination of $88\fdg01$ \citep{2007A&A...471..765B}, the galaxy is
displaying a dust lane, visible along the entire disc and with an
extensive vertical dust distribution
\citep{1999AJ....117.2077H,2004AJ....128..662T}, although we note that
\citet{2000A&AS..145...83A} consider this galaxy to be too highly
inclined to make any unambiguous claims about extraplanar dust. A
global fit of extinction by the dust distribution to optical images
was carried out in \citet{2007A&A...471..765B}, giving a dust
scalelength of 70\arcsec or 6.7~kpc, which is 75\% larger than the
stellar disc scalelength.

Compared to the other galaxies in the sample, the horizontal profiles
for NGC\,4217 are remarkably smooth: only the 100~$\mu$m\ profile --
and the 160~$\mu$m\ one to a lesser extent -- shows some small secondary peaks
at 0\farcm4 and 0\farcm6 from the centre, and two more, less prominent, at 1\arcmin and 1\farcm4, but all on the SW side. However, the overall
behaviour is quite peculiar and differs between wavelengths. At
100~$\mu$m, the profile first drops off from the centre outward, until
about 1\farcm6 on the SW side and about 1\arcmin on the NE side of the
disc. At these points, a break in behaviour occurs and the slope of
the profile becomes much steeper, up to about 2\arcmin on either side,
where the background takes over. For the other wavelengths, the
profiles have a slightly different character: on the SW side of the
disc, they run more or less parallel to the 100~$\mu$m\ profile until
the break in slope at about the same position, but beyond this point
they are not as steep as the 100~$\mu$m\ profile and reach out to
about 2\farcm5. On the other hand, at the NE end of the disc they seem
to have a single slope out to 2\farcm5.

The exponential model we superimposed to the observed horizontal
profile seems to be a good representation of the data, especially in
the innermost regions.  The PACS 100~$\mu$m\ map shows two clearly
visible secondary peaks, along with one (possibly two) fainter
structure(s), but due to the fact that all these are located on just
one side of the disc (the SW end), lacking any clearly distinguishable
counterparts on the other side, it would be very hard to identify any
of these as a ring or spiral arm with any degree of certainty. Due to
the worsening angular resolution, these structures can no longer be
told apart at the longer wavelengths.

\subsection{NGC\,5529}

\begin{figure*} \centering
  \includegraphics[width=\textwidth]{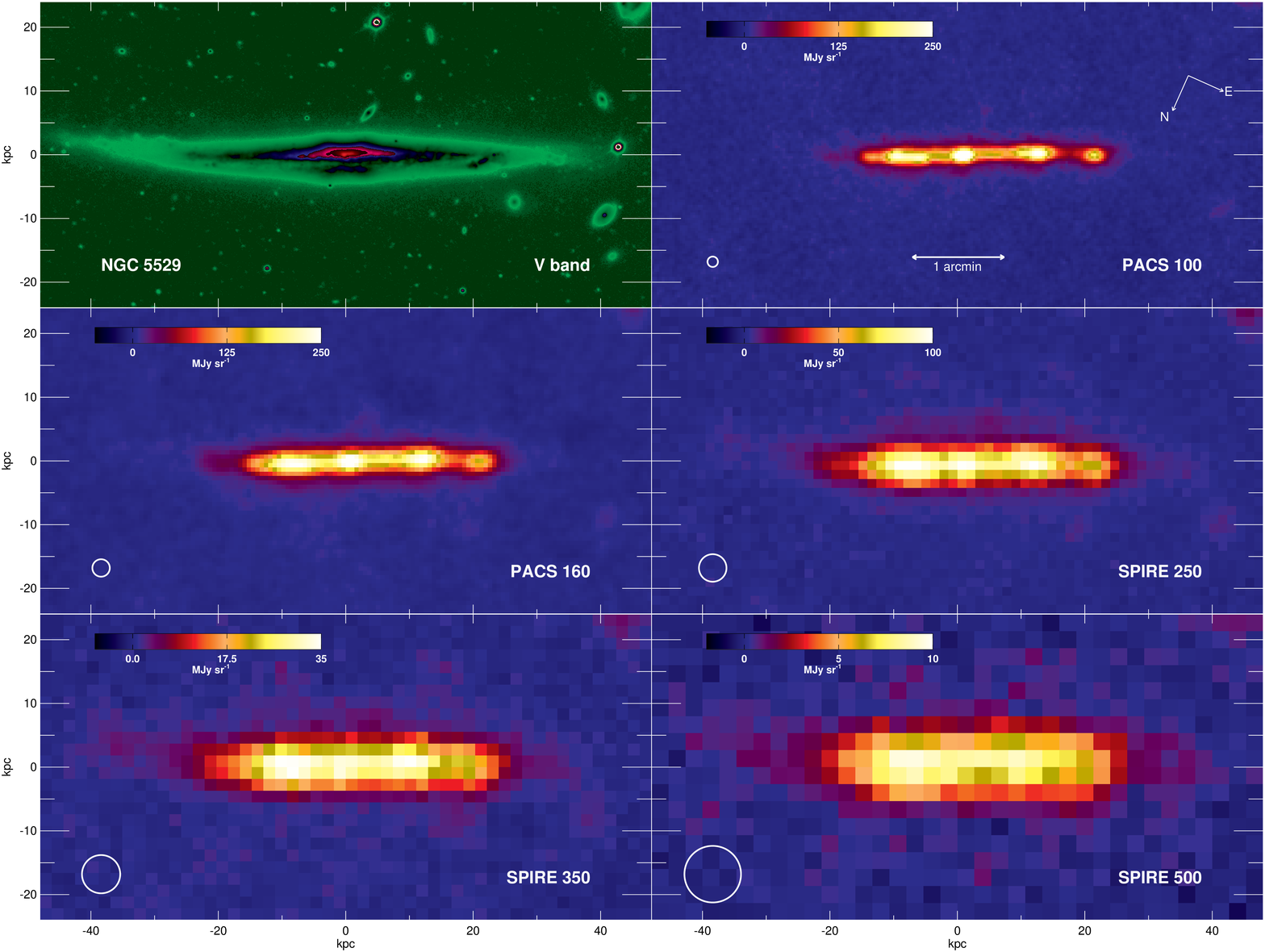}
  \caption{Optical V-band and \textit{Herschel} PACS and SPIRE images for
    NGC\,5529. The image sizes are about $6' \times 3'$.}
\label{NGC5529.fig}
\end{figure*}

With a distance of 49.5~Mpc, NGC\,5529 has a $D_{25}$ diameter of
6\farcm35 and is classified as an Sc type galaxy
\citep{1999A&A...344..868X}. The system shows a clear warp \citep[][in
POSS Blue and Red]{1990MNRAS.246..458S} and a box-shaped bulge
\citep{1987A&AS...70..465D}. The fitted inclination is $86\fdg94$
\citep{2007A&A...471..765B}, exposing a very prominent dust lane
\citep{1996A&AS..117...19D} along the major axis as well as in the
vertical direction \citep{1996A&AS..117...19D}. The dust distribution
was previously modelled in \citet{1999A&A...344..868X} and
\citet{2007A&A...471..765B} based on a global fit to optical
extinction data, both determining a dust scalelength of 50\arcsec,
corresponding to 11.7~kpc \citep{2007A&A...471..765B} or 11.9~kpc
\citep{1999A&A...344..868X}, which is about 35\%
\citep{2007A&A...471..765B}, respectively 50\%
\citep{1999A&A...344..868X}, larger than their fitted stellar disc
scalelength.

All PACS and the SPIRE 250~$\mu$m\ horizontal profiles have a clear
central peak, along with two secondary peaks on the SE side at
0\farcm8 and 1\farcm5 from the centre and another secondary peak
around 0\farcm6 on the other side, the latter having a small plateau
at PACS wavelengths. On the other hand, the other SPIRE horizontal
profiles show no apparent central peak and only marginally
differentiable peaks at the locations of the secondaries visible at
the previous wavelengths. At SPIRE wavelengths, the emission in the
central region remains more or less at a plateau out to around
1\farcm5 on the SE side and to around 0\farcm8 on the NW side, after
which the profiles drop off rather sharply out to 2\arcmin on both
sides, then reach another plateau at a reduced level, and finally fall
steeply around 2\farcm5 from the centre. For the PACS wavelengths, the
profile behaviour is quite different: while the NW side can be
interpreted as having a constant plateau underneath the secondary peak
out to 0\farcm8 from the centre, the SE side shows a shallow but clear
slope out to about 1\farcm6 from the centre. Beyond these edges, the
PACS profiles drop off sharply out to about 2\arcmin on either side.

A single dust disc with an exponential distribution does not suffice
to describe the FIR emission for this galaxy, as can be concluded from
the horizontal profiles, mostly due to the central regions where the
profiles are relatively flat. The mentioned secondary peaks can be
clearly distinguished in the images up to 250~$\mu$m, but given their
differing positions on either side of the central peak, one would tend
to conclude it is more likely that it concerns a spiral structure
rather than a few ring-like structures. A final remark on this galaxy
is the possibility of a warp in the dust disc: at least at 100~$\mu$m,
the outer secondary peaks (at 1\farcm4 on the SE side and at 0\farcm6
on the NW side) are collinear with the centre, but the brightest
secondary peak (at 0\farcm8 from the centre on the SE side of the
disc) seems to be concentrated slightly to the SW side of the major
axis. This could mean that this brightest secondary peak lies just
outside the main dust disc, but perhaps a more likely possibility is
that the whole dust disc has a position angle different from the
adopted V-band value, with the 0\farcm8 (SE) and 0\farcm6 (NW)
secondary peaks more or less collinear with the bulge, while the outer
peak at 1\farcm4 (SE) is warped outside of the dust disc. It should be
pointed out that \citet{2007A&A...471..765B} notes a slight warp in
the outer parts of the disc is clearly visible in the optical
observations.

\subsection{NGC\,5907}

\begin{figure*} \centering
  \includegraphics[width=\textwidth]{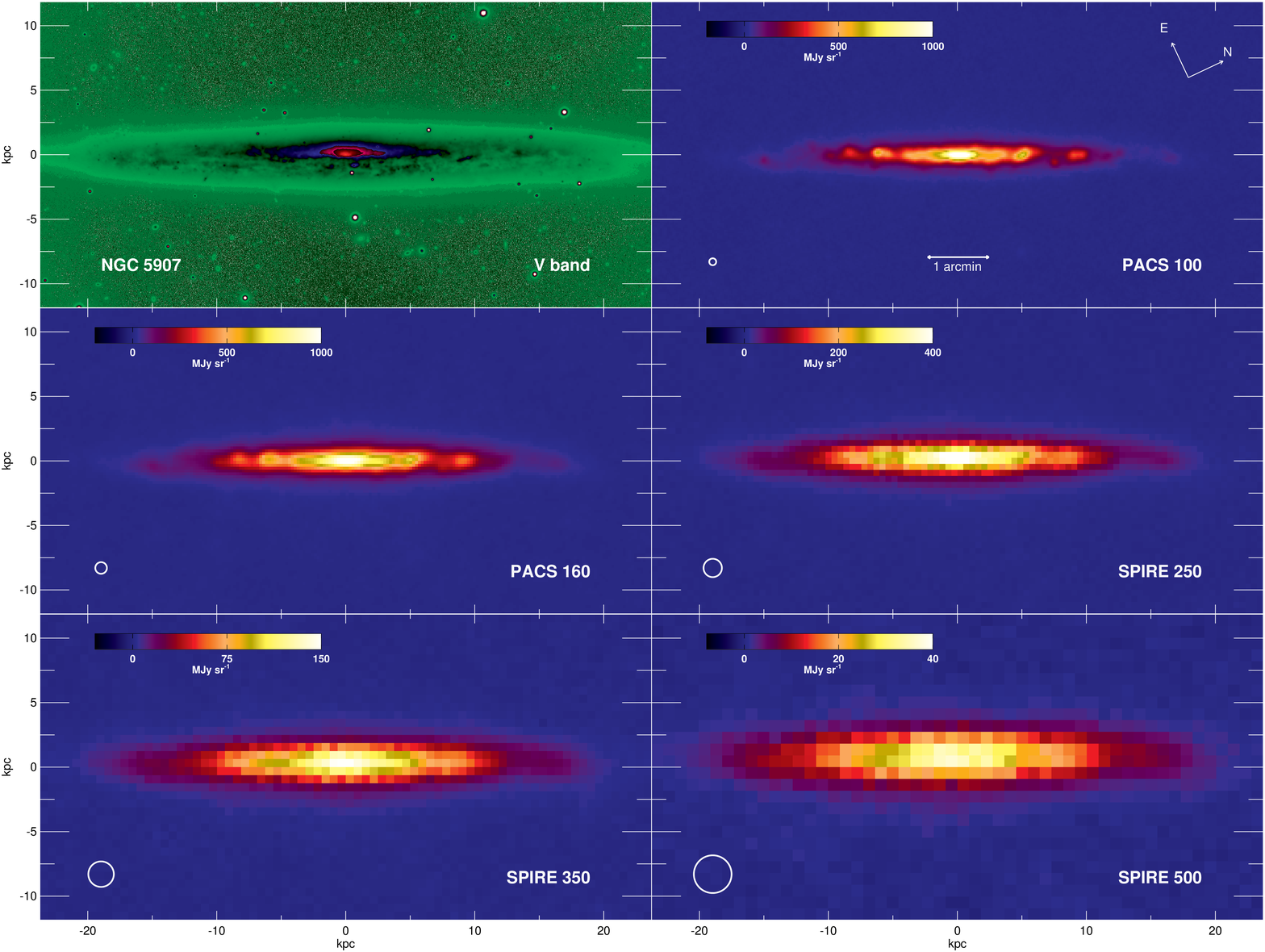}
  \caption{Optical V-band and \textit{Herschel} PACS and SPIRE images for
    NGC\,5907. The image sizes are $10' \times 5'$.}
\label{NGC5907.fig}
\end{figure*}

The galaxy NGC\,5907 is at a distance of 16.3~Mpc and has a $D_{25}$
diameter of 12\farcm77. It is considered to be either an Sc
\citep{1999A&A...344..868X} or an SA(s)c type galaxy and it has
H{\sc{ii}} regions. According to \citet[][in POSS Blue and
Red]{1990MNRAS.246..458S} it has a barely perceptible warp, but the
stellar \citep{1994AJ....108.1191M} and H{\sc{i}}
\citep{1976A&A....53..159S,1998ApJ...504L..23S} warps are definitely
both bending in the same direction. \citet{1994cag..book.....S} claim
the galaxy has a thin disc with the central bulge absent, but a bulge
was fitted well with a modified Hubble profile by
\citet{1992AJ....103...41B}, although resulting in a bulge-to-disc
luminosity ratio of only 0.05 in the H-band. The inclination for the
system has a fitted value of $87\fdg2\pm0\fdg2$
\citep{1999A&A...344..868X}. The dust lane extends from the centre out
to the edges along the major axis, and although
\citet{1999AJ....117.2077H} state that the galaxy has no detectable
vertically extended dust or ionised gas, \citet{2000A&AS..145...83A}
claim this galaxy to be insufficiently edge-on to make any definite
statements about extraplanar dust. The dust distribution has been
previously modelled using global fits to the dust lane extinction in
optical images \citep{1999A&A...344..868X}, leading to a dust
scalelength of 100\arcsec or 7.8~kpc, which is about 10\% larger than
the corresponding stellar disc scalelength.

Up to 250~$\mu$m, the horizontal profiles display a number of clear
secondary peaks on each side of the central peak, more or less
symmetrically positioned. All but the 100~$\mu$m\ profile have roughly
the same behaviour, with a single slope from the centre out to around
3\farcm5 from the centre and then a steeper slope out to about
5\arcmin to 6\arcmin on either side, although the 500~$\mu$m\ profile
shows a slightly shallower decline in the central area up to about
3\arcmin on both sides. The PACS 100~$\mu$m\ has a steeper drop-off
beyond 2\arcmin from the centre out to around 4\arcmin on either side,
where the background starts to dominate.

The exponential dust distribution is in this case a very good
description of the FIR horizontal profiles, at least at SPIRE bands,
also considering the disc is possibly truncated.  The fact that there
are several secondary peaks distinguishable at PACS wavelengths on
either side of the centre, could be indicative that a spiral structure
is more likely than a dust ring.

\begin{figure*} 
  \centering
  \includegraphics[width=0.32\textwidth]{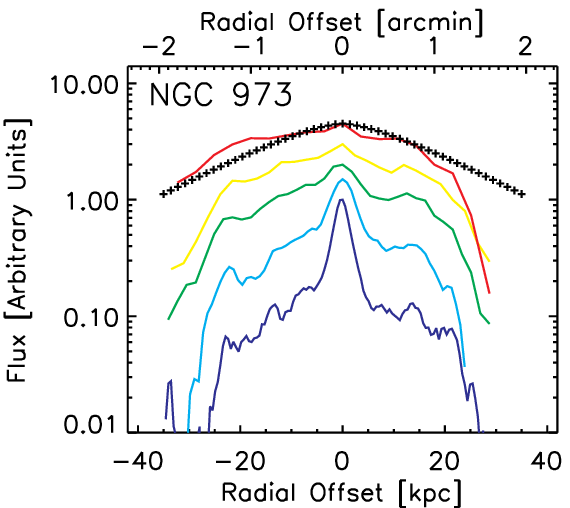} 
  \includegraphics[width=0.32\textwidth]{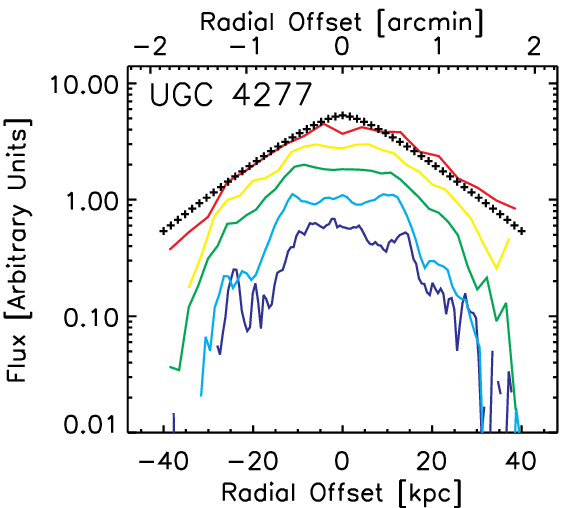}
  \includegraphics[width=0.32\textwidth]{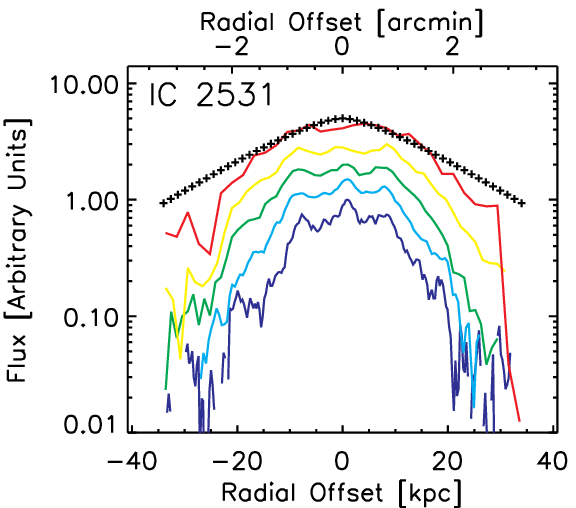} 
  \includegraphics[width=0.32\textwidth]{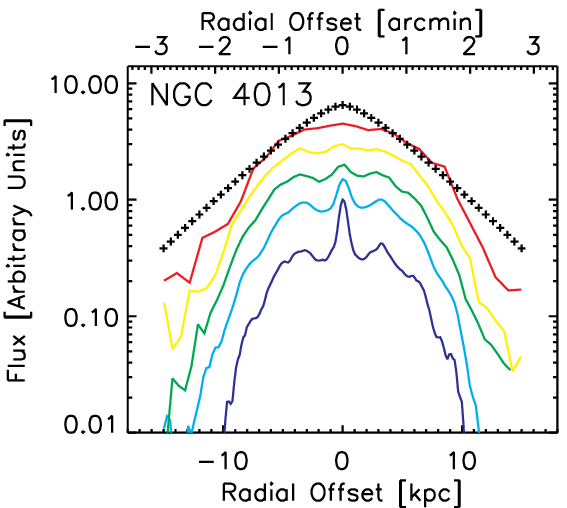}
  \includegraphics[width=0.32\textwidth]{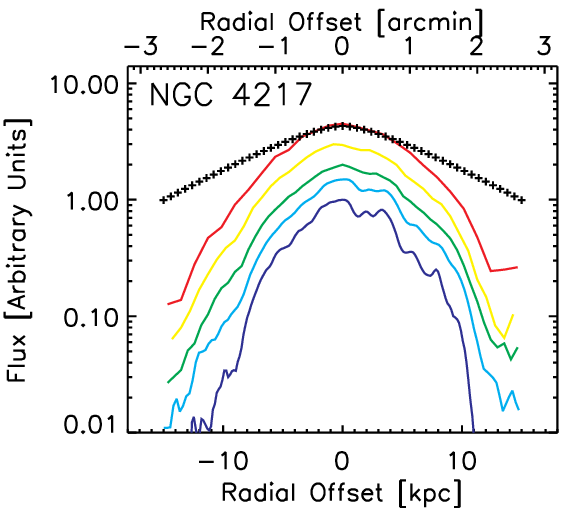}
  \includegraphics[width=0.32\textwidth]{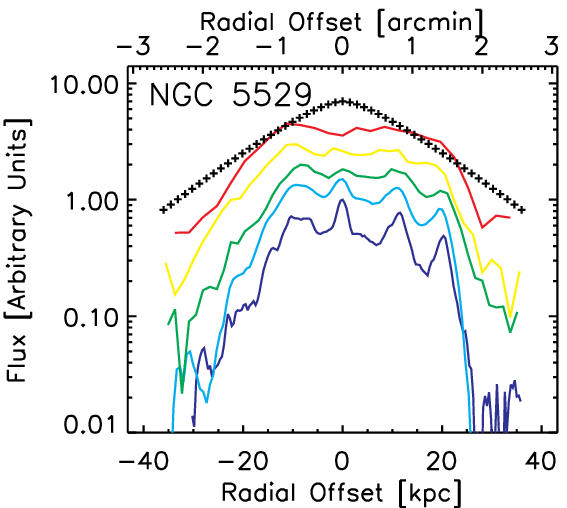}
  \includegraphics[width=0.32\textwidth]{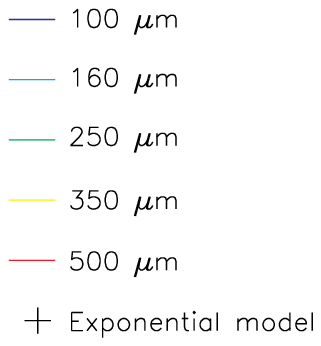}
  \includegraphics[width=0.32\textwidth]{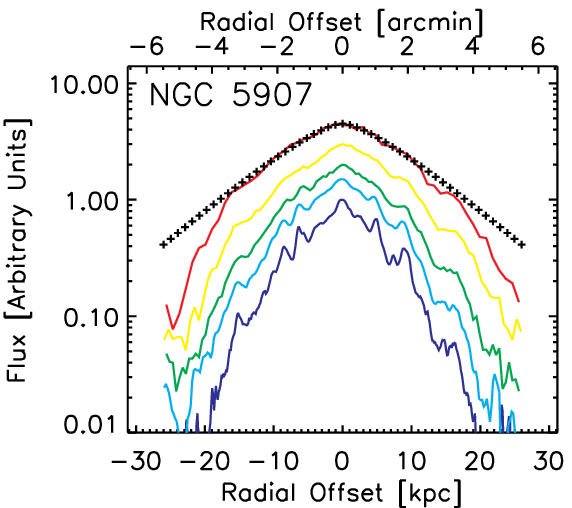}
  \hspace*{0.32\textwidth}
  \caption{The horizontal profiles for the galaxies in the sample, at
    \textit{Herschel} wavelengths. An arbitrary offset was
    introduced between different wavelengths for the sake of
    visualisation. A corresponding profile resulting from the
    double-exponential model is overplotted with black crosses, to
    compare with the commonly adopted description of the spatial dust
    distribution. The profiles are oriented as in the maps.}
\label{HorizontalProfiles.fig}
\end{figure*}

\section{Vertical dust distribution}
\label{VerticalProfiles.sec}

\begin{figure*} 
  \centering 
  \includegraphics[width=0.49\textwidth]{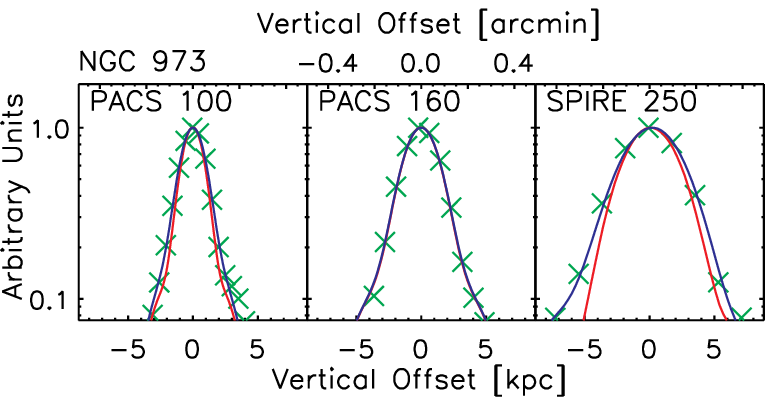}
  \includegraphics[width=0.49\textwidth]{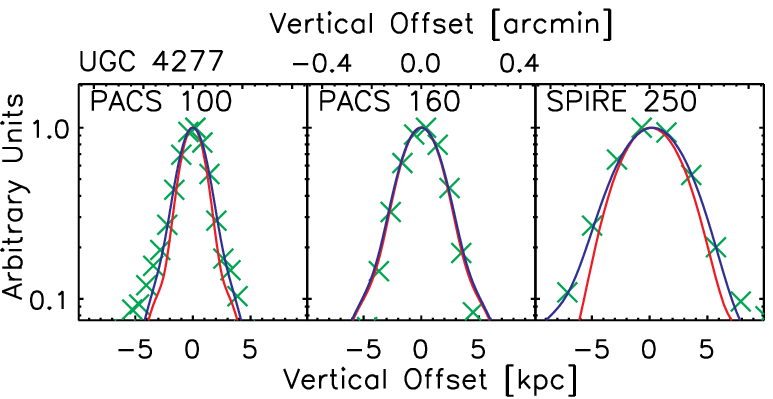}
  \includegraphics[width=0.49\textwidth]{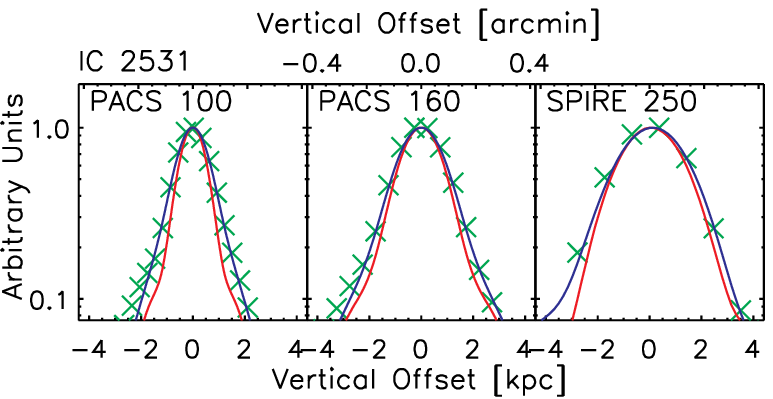}
  \includegraphics[width=0.49\textwidth]{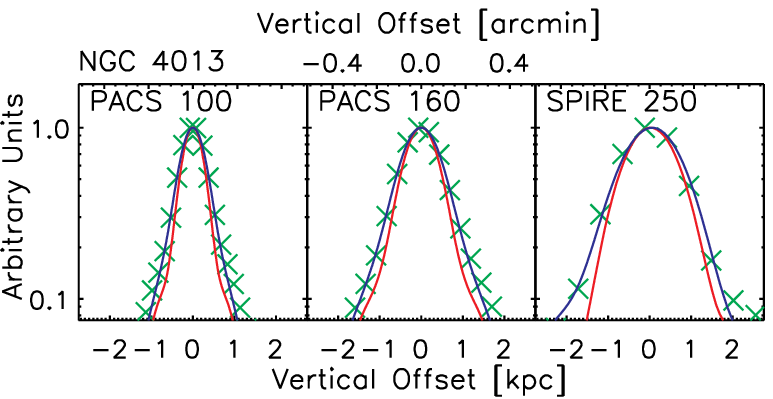}
  \includegraphics[width=0.49\textwidth]{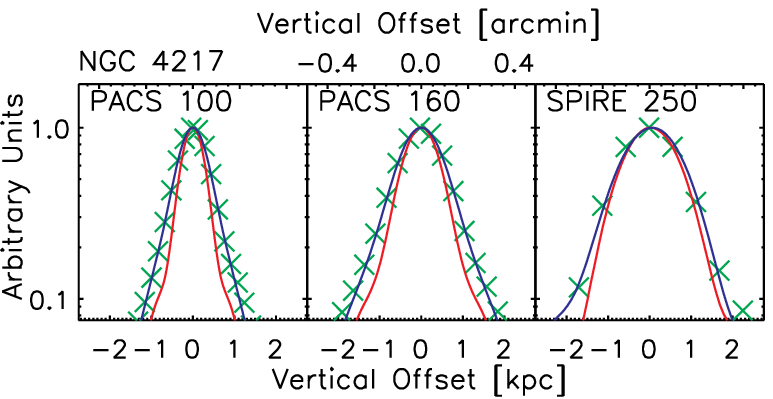}
  \includegraphics[width=0.49\textwidth]{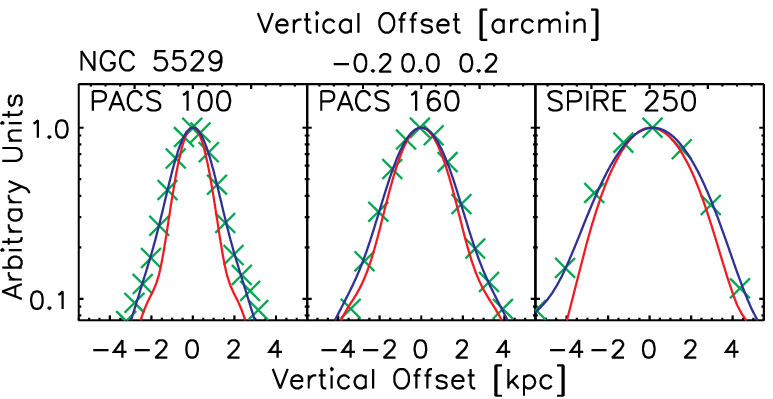}
  \includegraphics[width=0.49\textwidth]{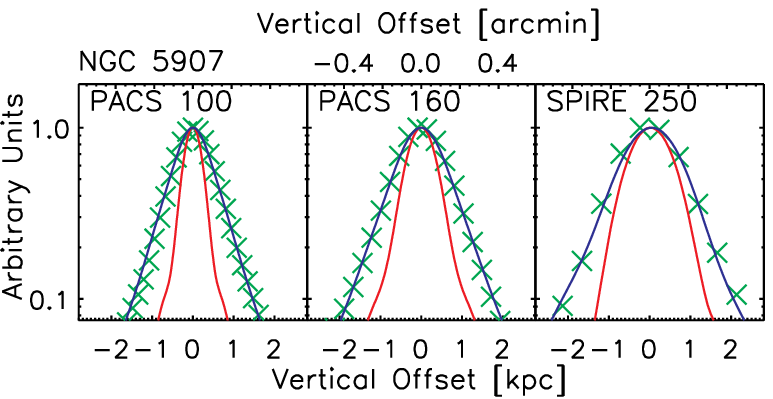}
  \caption{Vertical, normalised, profiles at 100, 160 and 250~$\mu$m\
    are plotted as green crosses, to which we superimpose, as a blue
    line, a fit with an exponential profile, convolved with the
    \textit{Herschel} beam at the appropriate wavelength (here
    depicted as a red line).}
  \label{VerticalProfiles.fig}
\end{figure*}
 
In this Section, we discuss the vertical distribution of the FIR/sub-mm
emission in the \textit{HER}OES galaxies. As mentioned in the
Introduction, edge-on spiral galaxies are the only systems where we
can directly infer the vertical distribution of the dust. The presence
of a spatially resolved vertical profile in the FIR for an edge-on
galaxy might be related to dust above the galactic disc (extra-planar
dust), ejected from the galactic plane.  Whether such ``dusty thick
discs'' are a common feature in spiral galaxies, and what the
correlation is with other physical properties, is not yet well
understood. Furthermore, studying the mechanisms that are responsible
for dust ejection can shed new light on the transport of ISM from the
galactic plane to higher latitudes.

Studying the vertical distribution of dust can be done in two ways:
either by analysing the absorption in optical images, or by directly
studying the vertical distribution of the FIR/sub-mm emission.
Radiative transfer modelling efforts of optical images of edge-on
spiral galaxies typically result in a dust disc with a vertical
scaleheight that is approximately only half the vertical scaleheight
of the stellar distribution, implying that the dust has a smaller
vertical velocity dispersion than the stars. Deep optical imaging,
however, has shown significant amounts of extraplanar dust in several
edge-on spirals \citep[e.g.][]{1999AJ....117.2077H,
  2000A&A...356..795A, 2004AJ....128..662T}. This dust at high
latitudes can be interpreted as either expelled outflowing material
from the galaxy if it is present, or infalling intergalactic matter
otherwise \citep{2009arXiv0904.4928H}. Small dust grains and
polycyclic aromatic hydrocarbons (PAHs), both relatively warm, have
been traced in direct emission up to several kpc above the disc plane
of some edge-on spirals \citep[e.g][]{2006A&A...445..123I,
  2007A&A...474..461I, 2007A&A...471L...1K, 2009MNRAS.395...97W}.

As already explained in Section \ref{DataReduction.subsec}, we have,
first of all, reprojected all the images so that the major axes of the
galaxies are parallel to the $x$ axes of the maps (see
Figure~\ref{NGC973.fig} to \ref{NGC5907.fig}). We subsequently
extracted vertical profiles at PACS and SPIRE wavelengths for each
galaxy. To do so, we first selected the area, along the horizontal
extent, where any pristine emission from the galaxy could be securely
detected. We then summed all the pixel values along the image $x$
coordinate (along the major axis) and normalised the resulting
profiles. The resulting vertical profiles are shown in
Figure~{\ref{VerticalProfiles.fig}}.

We modelled these vertical profiles with an exponential function of
the form:
\begin{equation}
  \label{eqn:prof}
  \Sigma_{\text{ver}}(z)
  =
  \frac{1}{2 h_z}
  \exp\left(-\frac{|z|}{h_z}\right)
\end{equation}
as appropriate for an exactly edge-on, double-exponential disc (see
Appendix~A). Before fitting this vertical profile to the observed
data, we first convolved it with the \textit{Herschel} beams at the
appropriate wavelength. The latter were obtained from the circularised
PSF images described in \citet{2011PASP..123.1218A}.  In order to
obtain the one-dimensional beams, we averaged the two-dimensional
PSFs along one direction, adopting a similar procedure as that
followed to produce the vertical profiles.  At this point we searched
the optimal value of $h_z$ that would best reproduce the observed
profile, by evaluating a standard $\chi^2$ function.
$N=9$ to 15 pixels were used to adequately sample the
observed vertical profiles, depending on the wavelength. Uncertainties
on the value of the scaleheight $h_z$ were also derived from the
$\chi^2$ probability distribution. The results of these fits are
listed in Table~{\ref{tab:hz}}, together with the value of the PSF in
linear units for each of the bands, and with the value of the vertical
scaleheight derived from radiative transfer modelling at optical
wavelengths. We consider that a galaxy has a spatially resolved
vertical profile when the profile is not dominated by the telescope
beam, that is when the deconvolved scaleheight value we derive from
the profile fitting is {\it not} consistent with zero at the 5$\sigma$
level.

Before the values of the scaleheight in different wavelengths can be
compared with each other, and with the scaleheight derived from
radiative transfer modelling, a number of issues need to be
considered.

A first obvious concern when comparing the dust scaleheights derived
in the different bands is the strongly different resolution of the
images. Even at the shortest wavelength we have observed, 100~$\mu$m,
the FWHM of the observations is $6\farcs8$ and it dominates the
vertical distribution. This is in vast contrast with the optical
images used for the radiative transfer modelling, which have
$\sim1\arcsec$ resolution. It can hence not be expected that both
methods give comparable results.

A second point to take into account is the effect of the
inclination. For the scaleheights derived from radiative transfer
modelling, the inclination of the dust disc has been taken into
account (the inclination is always one of the crucial free parameters
in the fitting procedure). In our simple fitting modelling approach of
the {\em{Herschel}} profiles, we implicitly assume a perfectly
edge-on disc. As the dust discs in spiral galaxies are extremely thin
and radially extended, even a small deviation from a perfect edge-on
orientation could strongly affect the observed ``vertical''
structure. In Appendix~A we calculate the apparent vertical
scaleheight that an infinitely thin exponential disc with radial
scalelength $h_R$ would appear to have, if it were projected on the
sky with inclination $i$,
\begin{equation}
  h_z 
  \approx
  1.8137\,h_R \cos i
\end{equation}
In the last column of Table~{\ref{tab:hz}} we list this apparent
scaleheight for the seven {\it{HER}}OES galaxies, based on the disc
scalelength and inclination derived from the radiative transfer
modelling.  By looking at the values reported in this table, together with the profiles in
Figure~{\ref{VerticalProfiles.fig}}, and taking into account the
cautions expressed above, we can draw different conclusions regarding the vertical structure of the
{\em{Herschel}} images.  For four out of the seven galaxies in the sample (NGC\,973, UGC\,4277,
IC\,2531 and NGC\,5529), the vertical profile is not resolved at the
5$\sigma$ level, even at the shortest wavelength. It is not a
coincidence that these four galaxies are the most distant ones in the
sample.

For two galaxies in the sample, NGC\,4217 and NGC\,5907, the vertical
profile is resolved at the 10$\sigma$ level, in the PACS 100 and
160~$\mu$m bands. Remarkably, in both cases the scaleheight of the
dust derived from the PACS observations is not in agreement with the
values derived from the radiative transfer fitting: in the case of
NGC\,4217 the FIR scaleheight is substantially smaller than the value
derived by \citet{2007A&A...471..765B}, whereas for NGC\,5907 it is
more than twice as large as the value obtained by
\citet{1999A&A...344..868X}. The key to these differences is the
inclination of the galaxies: they are both more than 2 degrees from
exactly edge-on. At these inclinations, the apparent vertical
structure of galaxies can be explained as an infinitely thin
exponential disc projected on the sky, as can be seen from the last
column of Table~{\ref{tab:hz}}.

Finally, for NGC\,4013, we resolve the vertical structure marginally
at the 5$\sigma$ limit in the PACS 100 and 160~$\mu$m bands. With an
inclination of 89\fdg7, this galaxy has the closest to exactly edge-on
orientation of all galaxies in the sample, and the vertical structure
cannot be due to the projection along the line-of-sight of the radial
structure. This is hence the only galaxy in the sample in which we
find reliable evidence for vertically resolved dust
emission. Interestingly, \citet{1999AJ....117.2077H} found evidence for
extra-planar dust from the analysis of extinction features in optical
images for NGC\,4013. The FIR-derived values for the scaleheight are
in good agreement with the value derived from radiative transfer
modelling by \citet{2007A&A...471..765B}. 

It is intriguing to see that
the scaleheight seems to increase with increasing wavelength. It is
tempting to interpret this as a natural consequence of the
decreasing dust temperature that is expected if one goes to gradually
higher distances above the plane of the galaxies. However, radiative transfer models have shown that
vertical gradients in T due to diffuse dust heating are very
shallow anyway \citep[see e.g. Figure~3 in][]{2000A&A...359...65B}.  Furthermore, this result
is most probably driven by the size of the PSF's FWHM which increases as a
function of the wavelength. Thus, it is normal that the best fit value
is somehow dependent on this (the same effect is also seen for the
galaxies where we do not properly resolve the vertical structure).

\begin{table*} 
  \caption{Scaleheight ($h_z$) values and related
    uncertainties as derived from the vertical profile fitting at 100,
    160 and 250 $\mu$m, together with the physical scale of the PSF,
    given at each wavelength.  The one but last column contains the vertical scaleheight 
    of the dust as derived from radiative transfer fits to
    V-band images by \citet{1997A&A...325..135X, 1999A&A...344..868X}
    or \citet{2007A&A...471..765B} (for the two galaxies in common between
    both samples, we list the average value). The last column contains
    the scaleheight that would result from an infinitesimally thin
    expontial disc, observed with the actual inclination of the
    galaxy (see Appendix). All scaleheights 
    and FWHM values are expressed in kpc.}
  \centering
  \begin{tabular}{lcccccccc}
    \hline
    \hline
    galaxy & $h_z [100]$ &  FWHM  & $h_z [160]$ &  FWHM  & $h_z [250]$
    &  FWHM  & $h_z[{\text{opt}}]$ & $h_z[{\text{incl}}]$ \\[+0.1cm]
    \hline
    \rule{0pt}{3.1ex}
    NGC\,973    & $0.36^{+0.16}_{-0.21}$ &   2.09  &
    $0.14^{+0.38}_{-0.13}$ &   3.41   &   $0.94^{+0.65}_{-0.92}$ &
    5.41   & $0.59$  & 0.21 \\[+0.1cm]
    UGC\,4277  & $0.48^{+0.36}_{-0.47}$ &   2.51  &
    $0.30^{+0.30}_{-0.30}$ &   4.08   &  $1.09^{+1.02}_{-1.07}$ &
    6.48  & $0.25$  & 0.44 \\[+0.1cm]
    IC\,2531       & $0.33^{+0.11}_{-0.11}$ &  1.23   &
    $0.33^{+0.15}_{-0.24}$ &   2.00   &  $0.44^{+0.37}_{-0.43}$ &
    3.17  & $0.38$ & 0.17 \\[+0.1cm]
    NGC\,4013  & $0.14^{+0.03}_{-0.03}$ &  0.62   &
    $0.21^{+0.05}_{-0.05}$ &   1.02   &  $0.29^{+0.11}_{-0.14}$ &
    1.61  & $0.20$ & 0.03 \\[+0.1cm]
    NGC\,4217  & $0.21^{+0.02}_{-0.02}$ &  0.66   &
    $0.27^{+0.04}_{-0.04}$ &   1.07   &  $0.26^{+0.10}_{-0.12}$ &
    1.70  & $0.38$ & 0.43 \\[+0.1cm]
    NGC\,5529  & $0.46^{+0.10}_{-0.10}$ &  1.64   &
    $0.42^{+0.17}_{-0.23}$ &   2.67   &  $0.74^{+0.40}_{-0.72}$ &
    4.24  & $0.39$ & 1.16 \\[+0.1cm]
    NGC\,5907  & $0.41^{+0.04}_{-0.04}$ &  0.56   &
    $0.42^{+0.04}_{-0.04}$ &   0.90   &  $0.45^{+0.10}_{-0.10}$ &
    1.43  & $0.16$ & 0.69 \\[+0.1cm]
    \hline
  \end{tabular}
  \label{tab:hz}
\end{table*}

\section{Determination of the dust masses}
\label{DustMasses.sec}

To determine the total mass of the dust in the galaxies, we follow two
different methods. Firstly, we use the results from fitting dust
distribution models to V-band images due to extinction as given in
\citet{1997A&A...325..135X,1999A&A...344..868X} and
\citet{2007A&A...471..765B}.  For the second method, we determine the
dust mass by fitting simple modified black-body models to the global
\textit{Herschel} fluxes.

\subsection{Dust masses from radiative transfer fits}

\begin{table*} 
  \centering 
  \caption{Derivation of the dust masses for
    the galaxies in our sample. The second, third and fourth columns
    contain horizontal scalelength $h_R$, the vertical
    scaleheight $h_z$, and the face-on optical depth
    $\tau_{\text{V}}^{\text{f}}$ of the dust as derived from radiative
    transfer fits to V-band images (the scalelength and scaleheights
    have been rescaled to the distances adopted within \textit{HER}OES). The
    fifth column is the optically determined dust mass calculated from these values
    using equation~(\ref{Mdopt}). The sixth column contains the
    reference of the radiative transfer fit: X97
    \citep{1997A&A...325..135X}, X99 \citep{1999A&A...344..868X} or B07
    \citep{2007A&A...471..765B}. Dust temperature and dust mass as derived
    from a modified black-body fit to the {\it Herschel} data are
    reported in the following columns, while the last one is the far
    infrared luminosity.}
  \label{DustMasses.tab}
  \begin{tabular}{lcccccccc}
  \hline \hline
  galaxy & $h_R$ & $h_z$ &
  $\tau_{\text{V}}^{\text{f}}$ & $\log M_{\text{d}}^{\text{opt}}$ & ref & $T_{\text{d}}$
  & $\log M_{\text{d}}^{\text{FIR}}$ \\
  & (kpc) & (kpc) & & ($M_\odot$) & & (K) & ($M_\odot$) \\
  \hline
NGC\,973   & 16.33 & 0.59 & 0.48 & 8.17 & X97   &   $20.0\pm0.6$ & $8.11\pm0.07$ \\
UGC\,4277 & 12.52 & 0.25 & 0.49 & 7.95 & B07   &  $17.3\pm0.6$ & $8.31\pm0.09$ \\
IC\,2531      & 13.68 & 0.38 & 0.30 & 7.81 & X99   &  $18.5\pm0.3$ & $8.04\pm0.04$ \\  
NGC\,4013 &   3.93 & 0.21 & 0.67 & 7.08  & X99  &  $21.5\pm0.4$ & $7.63\pm0.04$ \\
                     &   2.67 & 0.19 & 1.46 & 7.08 & B07   & & & \\
NGC\,4217 &   6.72 & 0.38 & 1.26 & 7.81 & B07    &  $22.1\pm0.4$ & $7.85\pm0.03$ \\
NGC\,5529 & 11.87 & 0.52 & 0.65 & 8.02 & X99   &   $19.4\pm0.4$ & $8.32\pm0.05$  \\
                     & 11.66 & 0.26 & 0.68 & 8.03 & B07   &  & & \\
NGC\,5907 & 7.84   & 0.16 & 0.49 & 7.54 & X99   &   $20.0\pm0.3$ & $8.12\pm0.03$ \\
\hline
\end{tabular}
\end{table*}

\citet{1997A&A...325..135X,1999A&A...344..868X} and
\citet{2007A&A...471..765B} determined the intrinsic distribution of
stars and dust in all \textit{HER}OES galaxies by fitting radiative
transfer models to V-band images. In their models, the dust mass
density distributions are smooth, axisymmetric models with the
double-exponential behaviour defined earlier in equation
(\ref{ded}).  Rather than the total dust mass, the
  quantity of dust in these models is usually parametrised by the
  face-on optical depth along the central line-of-sight. The
  connection between dust mass and the face-on optical depth follows
  directly from the definition of the latter quantity,
\begin{equation} 
  \tau_\lambda^{\text{f}} 
  = 
  \int_{-\infty}^\infty \kappa_\lambda\, \rho(0,z)\, {\text{d}}z
  \label{face-on-optical-depth}
\end{equation} 
with $\kappa_\lambda$ the extinction coefficient for the dust at
wavelength $\lambda$. Combining this expression with the density
distribution (\ref{ded}) of the double-exponential disc model gives
us
\begin{equation} 
  M_{\text{d}}
  = 
  \frac{2\pi\,\tau_\lambda^{\text{f}}\,
    h_R^{2}}{\kappa_\lambda} 
  \label{total-dust-mass-simplified}
\end{equation} 
Adopting the value $\kappa_{\text{V}} = 2619$~m$^{2}$~kg$^{-1}$ for
the V-band dust extinction coefficient in the interstellar medium
\citep{2003ApJ...598.1017D} leads to the formula
\begin{equation}
  M_{\text{d}}
  = 
  1.148 \times 10^6 \; \tau_{\text{V}}^{\text{f}}
  \left(\frac{h_R}{\text{kpc}}\right)^2 \; 
  M_{\sun}
\label{Mdopt}
\end{equation}
which is virtually identical to equation (11) from
\citet{1997A&A...325..135X}. When we substitute the values for $h_R$
and $\tau_{\text{V}}^{\text{f}}$ as derived by
\citet{1997A&A...325..135X, 1999A&A...344..868X} and
\citet{2007A&A...471..765B}, we obtain the dust masses listed in the
fifth column of Table~{\ref{DustMasses.tab}}, where we added the
superscript ``opt'' to indicate that it refers to the dust mass
determined from radiative transfer fits to optical images. For the two
galaxies in common between the samples used in the
  aforementioned works, NGC\,4013 and NGC\,5529, the derived dust
mass estimates are nearly identical. This is particularly remarkable
for NGC\,4013, since the values for $\tau_{\text{V}}^{\text{f}}$ and
$h_R$ obtained by the radiative transfer fits are quite
different.  This is due to a degeneracy in the radiative
  transfer modelling of edge-on spiral galaxies, which has been noted
  by \citet{2007A&A...471..765B} and
  \citet{2013A&A...550A..74D}. Systems with a large face-on optical
  depth and a small dust scalelength and systems with a small face-on
  optical depth and a large dust scalelength can both result in
  similar edge-on optical depth and hence dust lanes of similar
  depths. Fortunately, the total dust mass, which is a combination of
  these parameters, is rather insensitive to this degeneracy.

\subsection{Global SED fitting}
\label{ssect:sed}

\begin{figure*} 
  \centering
  \includegraphics[width=0.32\textwidth]{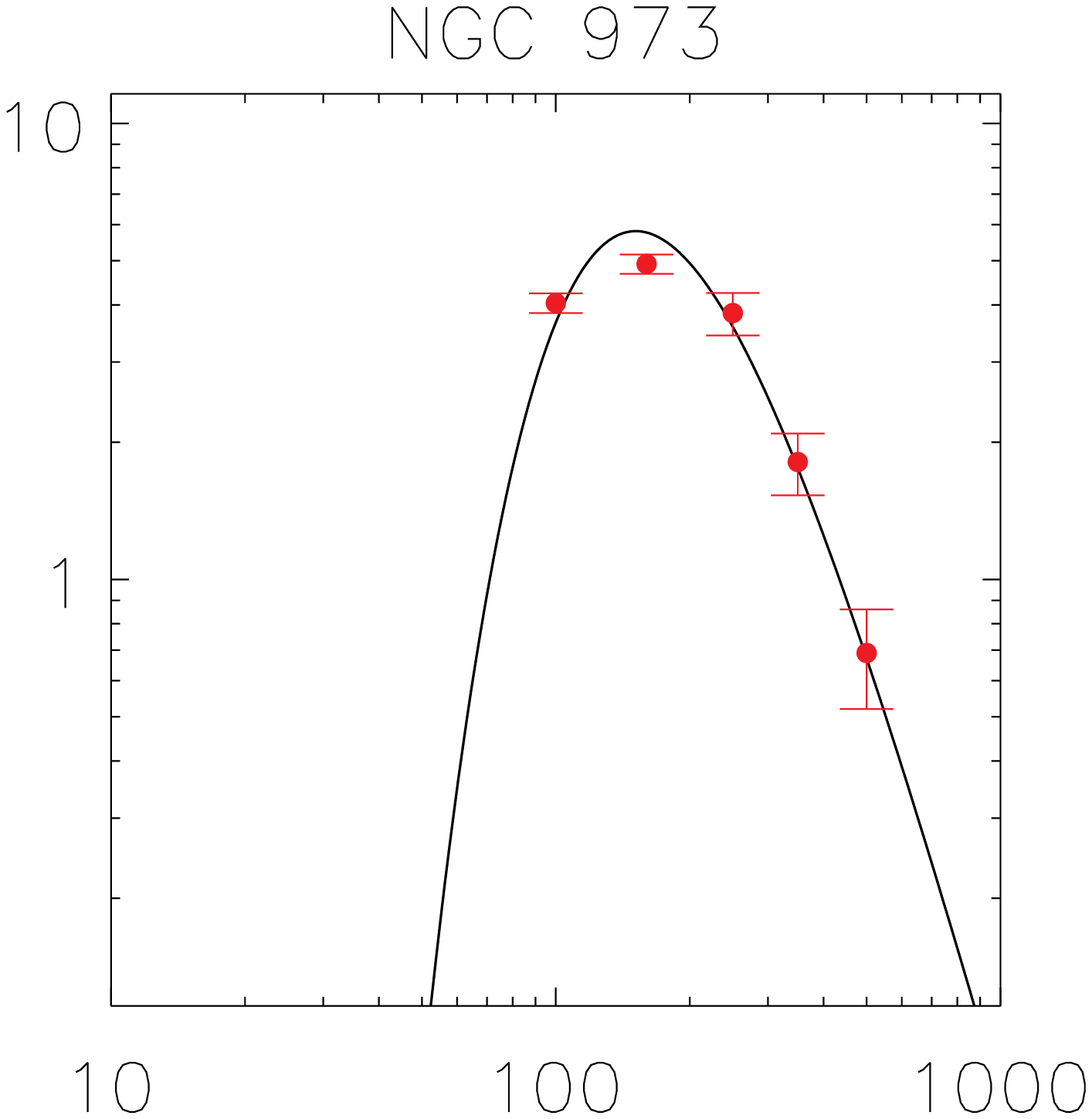}
  \includegraphics[width=0.32\textwidth]{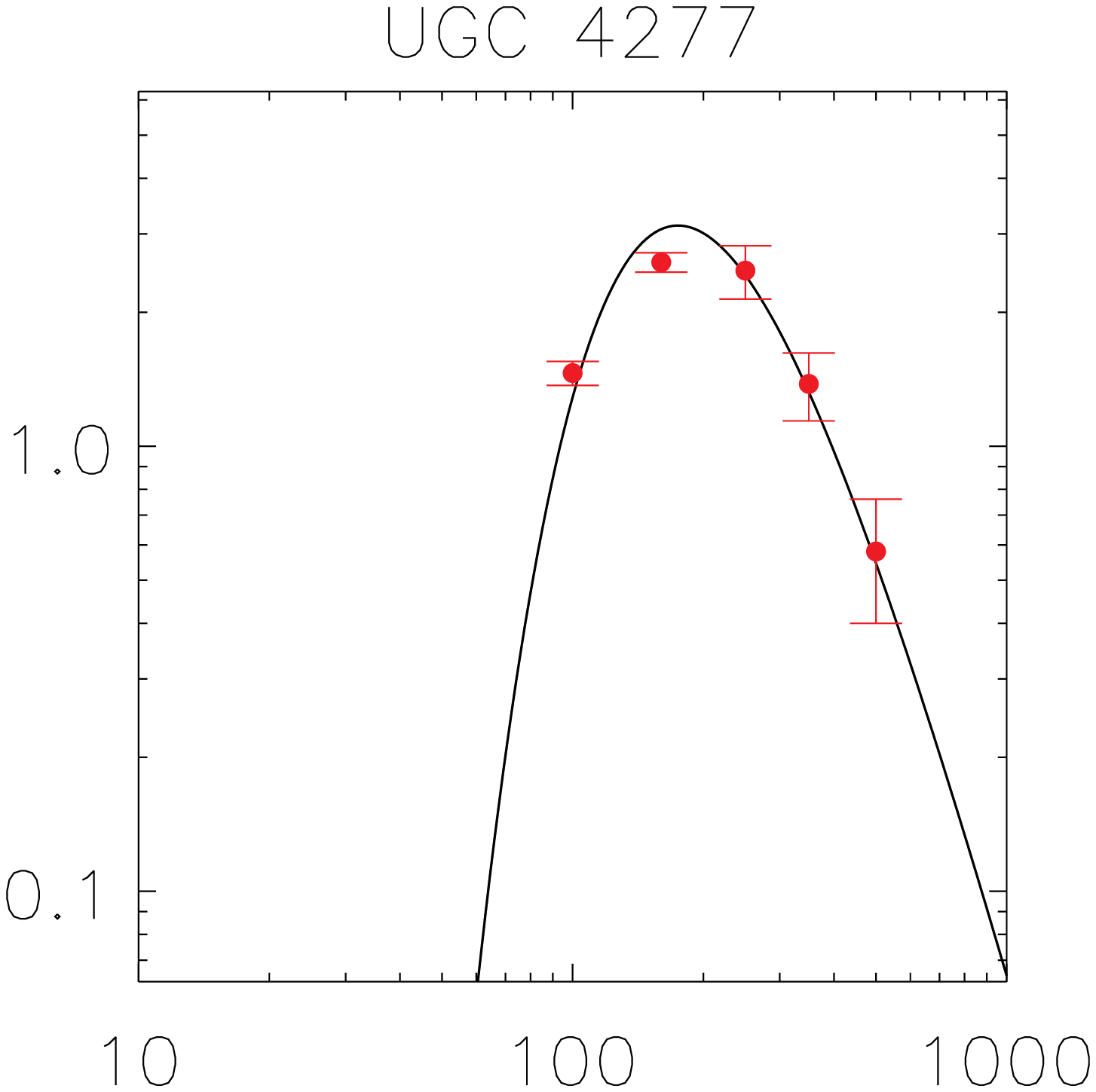}
  \includegraphics[width=0.32\textwidth]{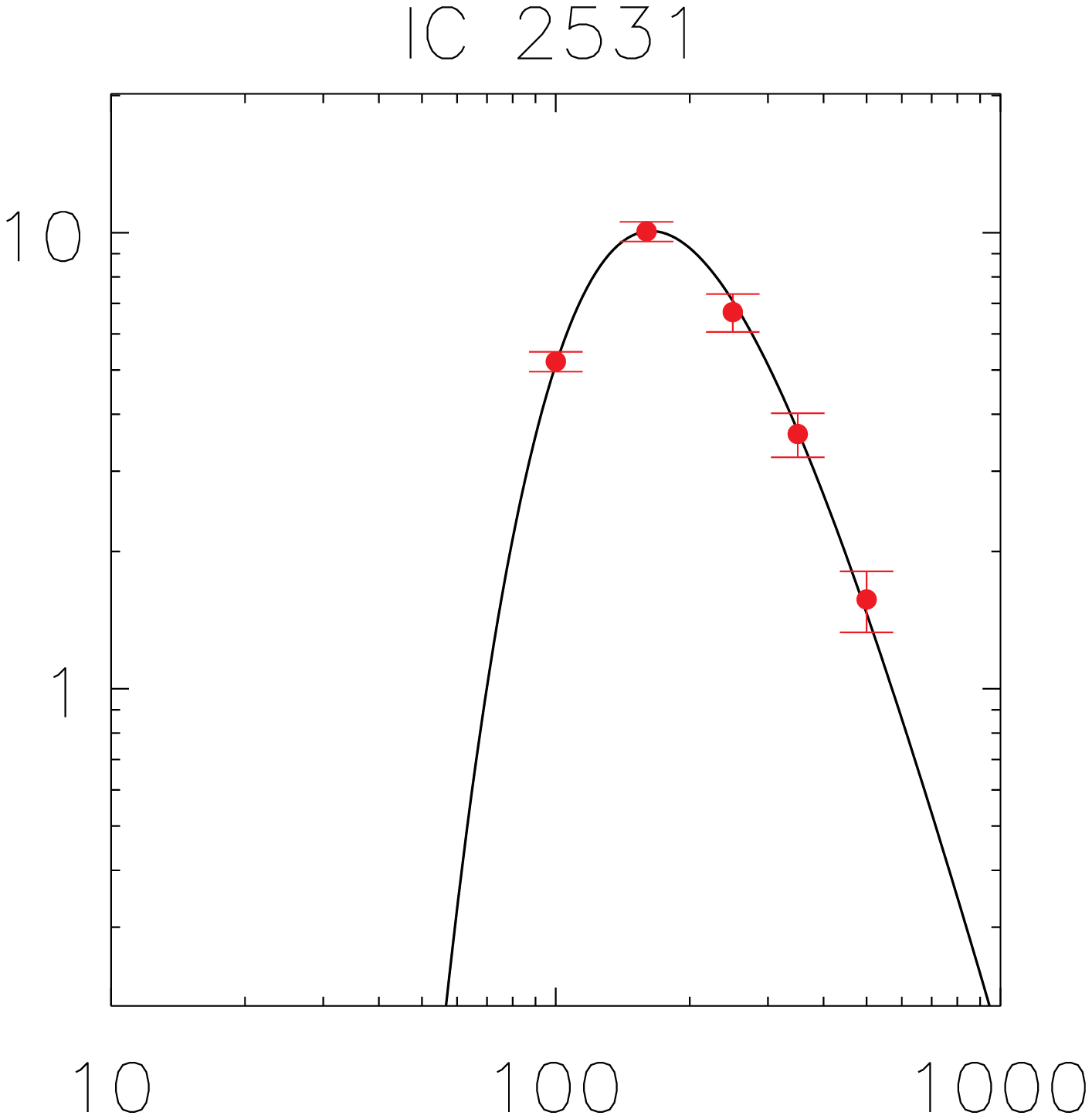} 
  \includegraphics[width=0.32\textwidth]{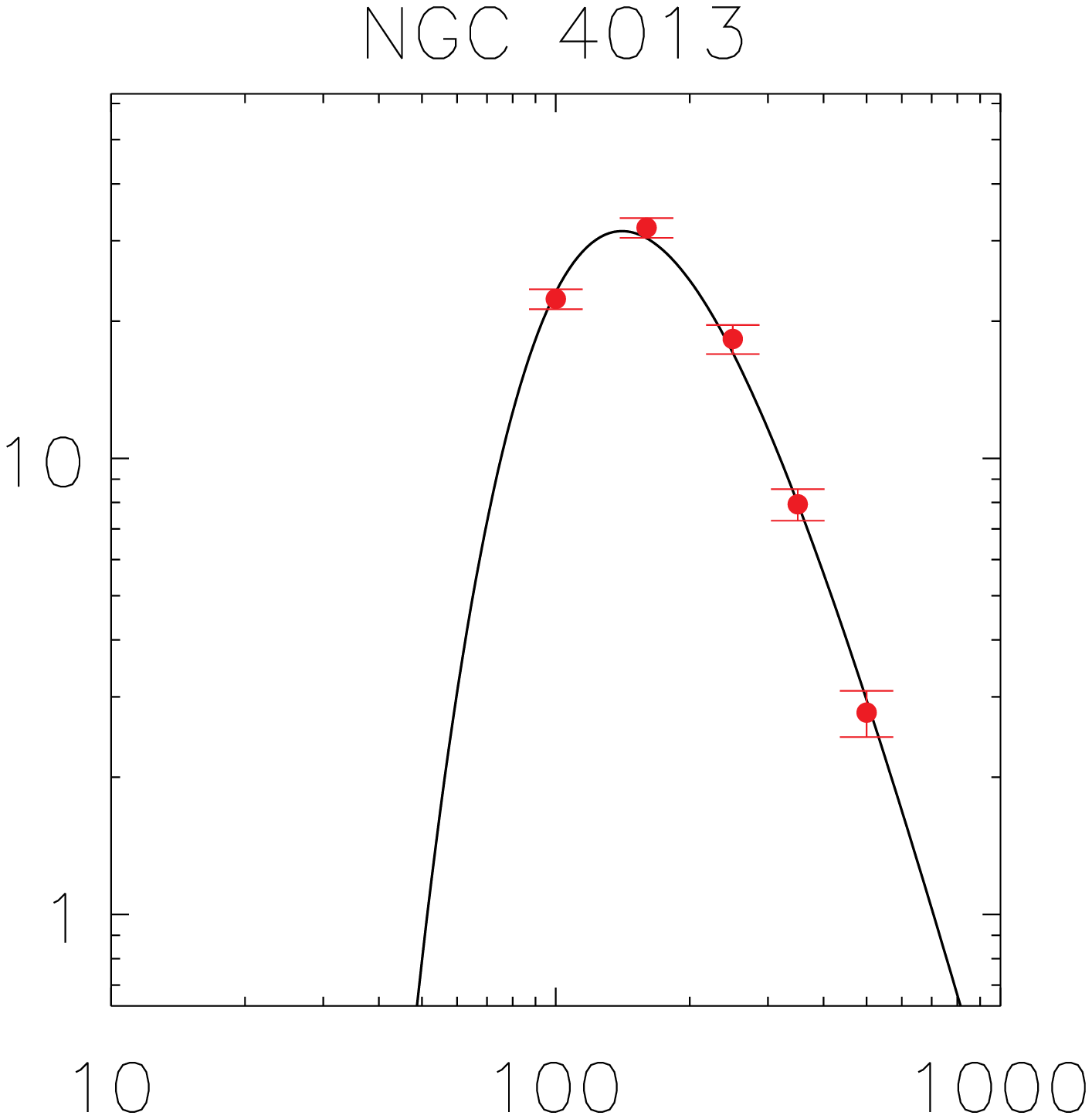}
  \includegraphics[width=0.32\textwidth]{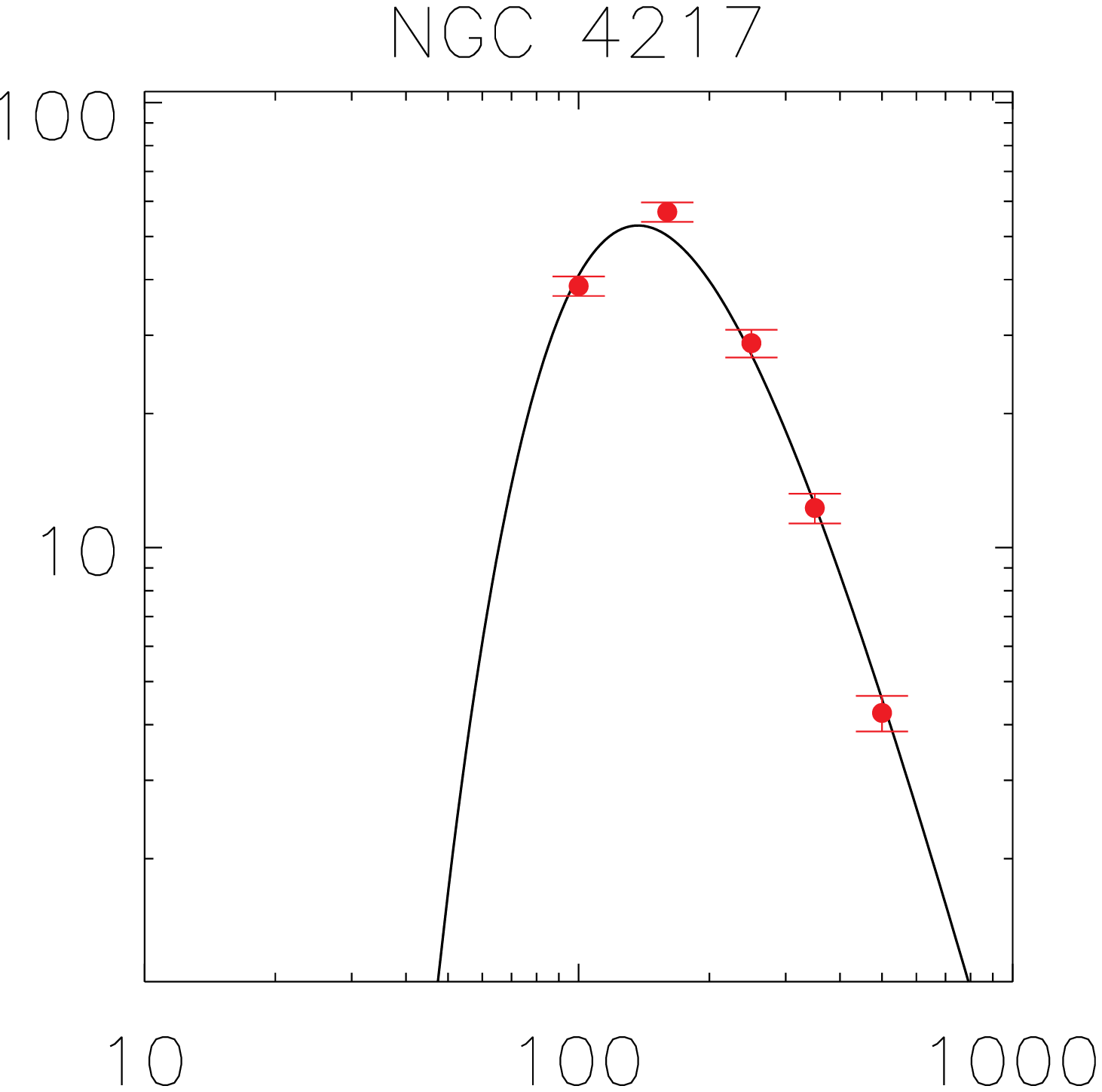}
  \includegraphics[width=0.32\textwidth]{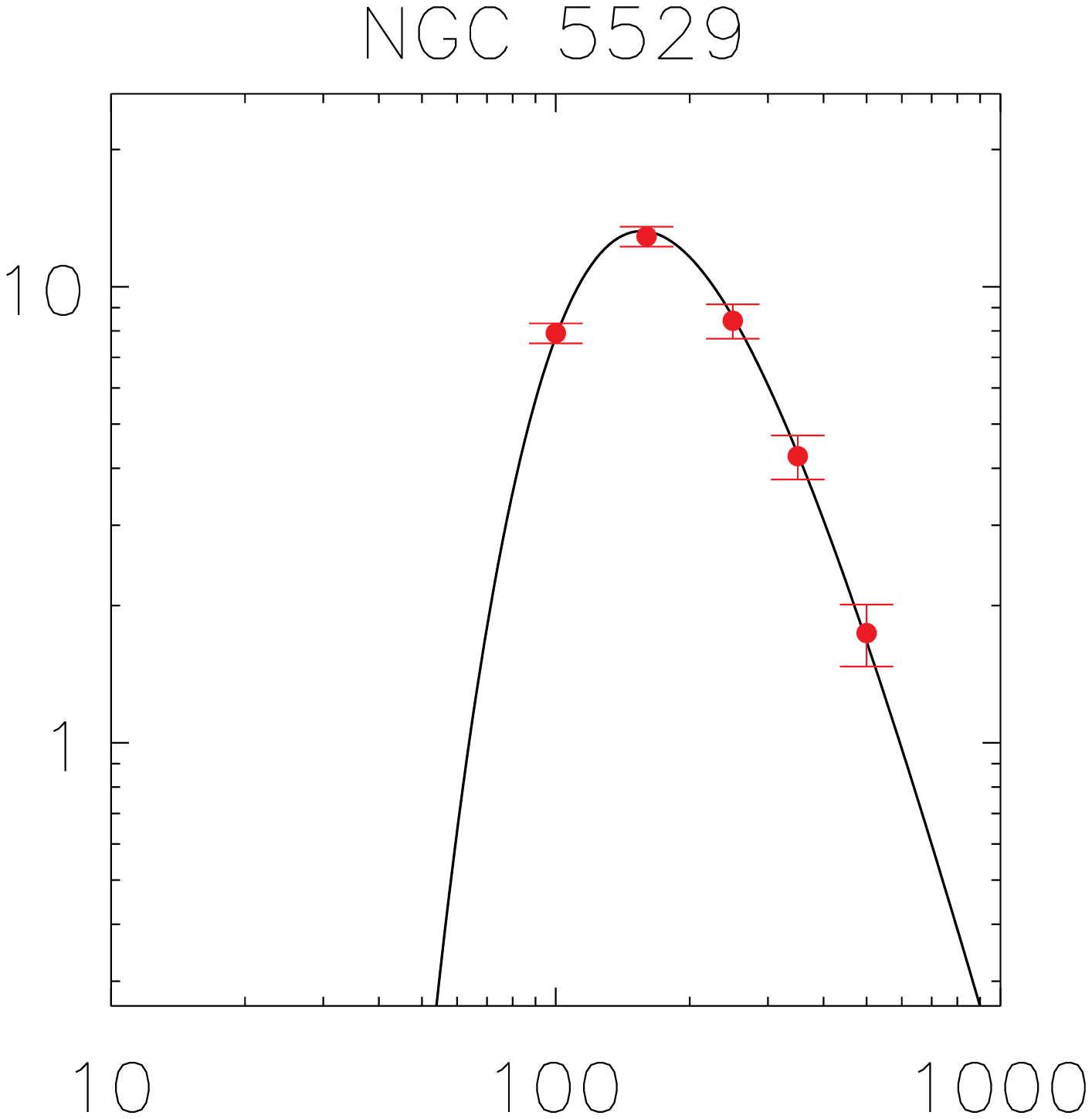}
  \includegraphics[width=0.32\textwidth]{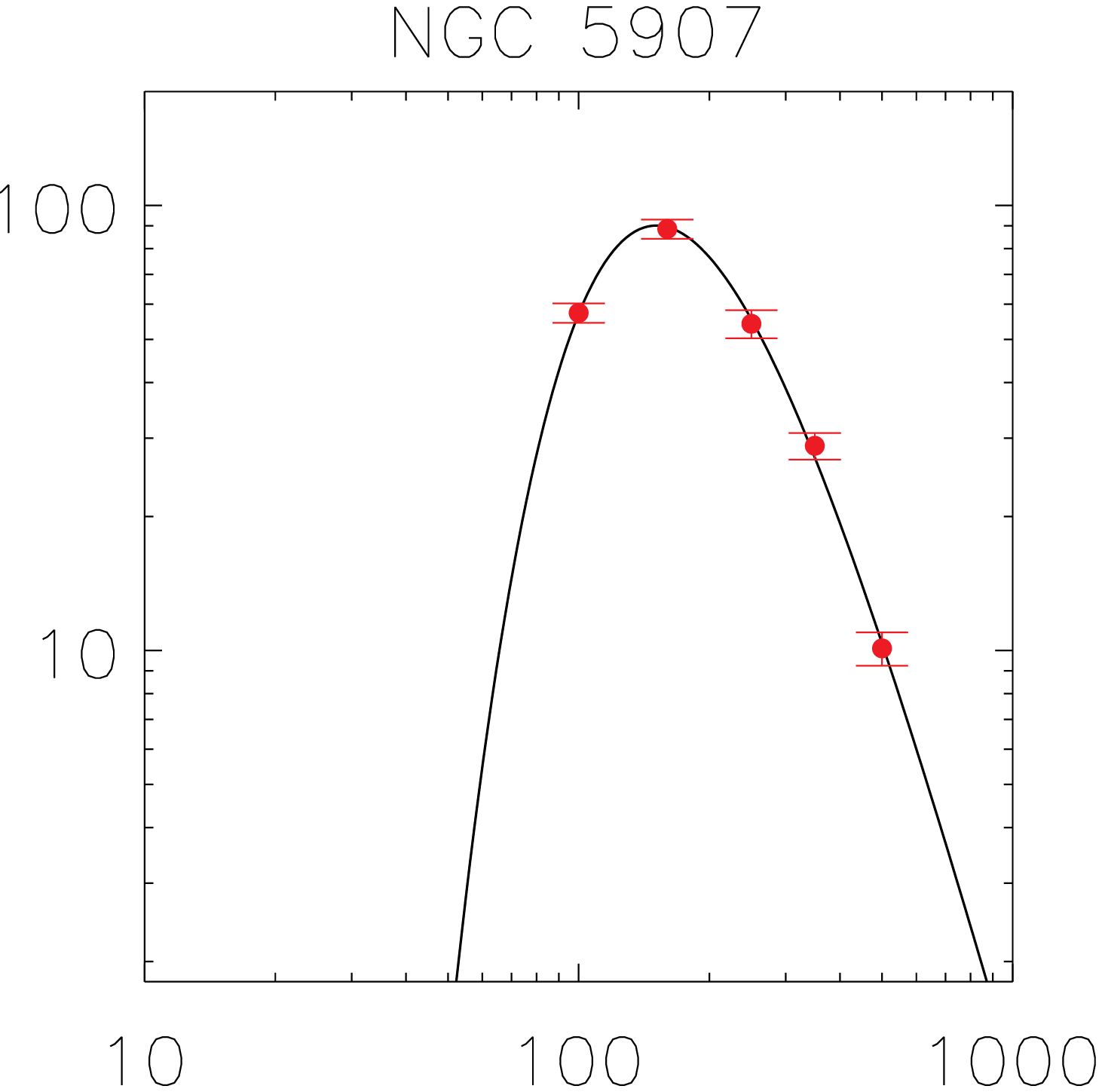}
  \caption{SED fitting to the \textit{Herschel} fluxes (here
    depicted as red diamonds) for all \textit{HER}OES
    galaxies. A modified, single temperature black-body model (black
    line) is adopted.  The y axes are in units of Jansky while the x
    axes are in micron.}
  \label{ModBB.fig}
\end{figure*}

We also determined the dust masses for the galaxies in our sample by
fitting a simple modified black-body model to the PACS and SPIRE data,
i.e.\ 
\begin{equation}
  F_\nu
  =
  \frac{M_{\text{d}}\,\kappa_\nu\, B_\nu(T_{\text{d}})}{D^2}
\end{equation}
where $M_{\text{d}}$ is the dust mass, $B_\nu$ is the Planck function,
$T_{\text{d}}$ is the dust temperature, $D$ is the distance to the
galaxy and $\kappa_\nu$ is the dust emissivity. As customary, we
assumed a power-law dust emissivity in the FIR/sub-mm wavelength range,
\begin{equation}
  \kappa_\nu \propto \nu^\beta
\end{equation}
and we fix the value of the emissivity to $\kappa_\nu = 0.192$ m$^2$
kg$^{-1}$ at 350 $\mu$m. The fits were done by performing a $\chi^2$
minimisation using a simple gradient search method, with
$M_{\text{d}}$ and $T_{\text{d}}$ as free parameters, with $\beta$
fixed to a value of 1.8. Error bars on the fitted parameters were
derived using a bootstrapping method, as follows: when
  the best--fit parameters ($T_{\text{d}}$, $M_{\text{d}}$ and
  $\beta$) are found, 200 new sets of datapoints are created by
  randomly drawing a value from the observed fluxes, lying within the
  observed errorbars. A best fit is searched for each one of these new
  artificial datasets: 16\% of the best fit models having the lower
  and higher parameters values are discarded, and uncertainties are
  then taken as the differences between the best fit solution and the
  extreme values.

The results of these modified black-body fits are shown in
Figure~{\ref{ModBB.fig}} and listed in the last three columns of
Table~{\ref{DustMasses.tab}} (the dust masses as derived from the SED
fits are denoted as $M_{\text{d}}^{\text{FIR}}$ to make the
distinction with the optically determined dust masses).  Had we left
the emissivity index, $\beta$, as a free parameter, we would have
found an average value of $\langle\beta\rangle = 1.73\pm0.36$, very
close to the value we have used for the fits.  Using the same value
for the emissivity index allows us to compare, in a consistent manner,
the dust mass values we derive. Furthermore, this value of $\beta$ and
the average temperature $\langle T_{\text{d}}\rangle = (19.8\pm1.6)$
K, that we obtain, are very typical values for the interstellar dust
medium in spiral galaxies \citep[e.g][]{2011MNRAS.417.1510D,
  2012MNRAS.419.3505D, 2012ApJ...756...40S, 2012A&A...540A..54B,
  2012MNRAS.425..763G}.  As the latter model only reproduces
datapoints longwards of 100~$\mu$m, the dust mass values we give do
not take into account the possible presence of warmer dust, whose
signature would show up at shorter wavelengths. The latter, anyway,
constitutes a minor fraction of the total dust mass \citep[see,
e.g.][]{2010A&A...518L..51S}.

\subsection{Comparison of optical and FIR dust masses}

\begin{figure*}
  \centering
  \includegraphics[width=\columnwidth]{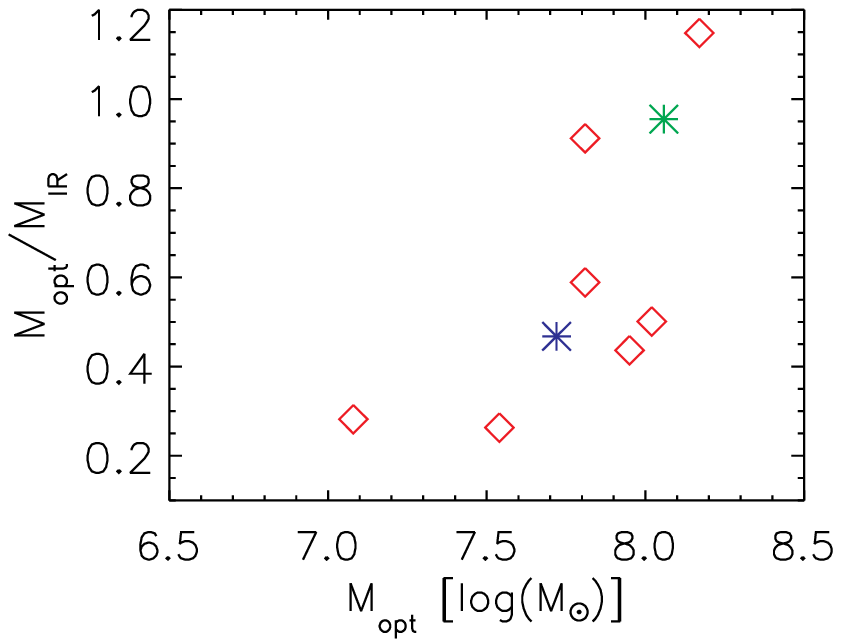} 
  \includegraphics[width=\columnwidth]{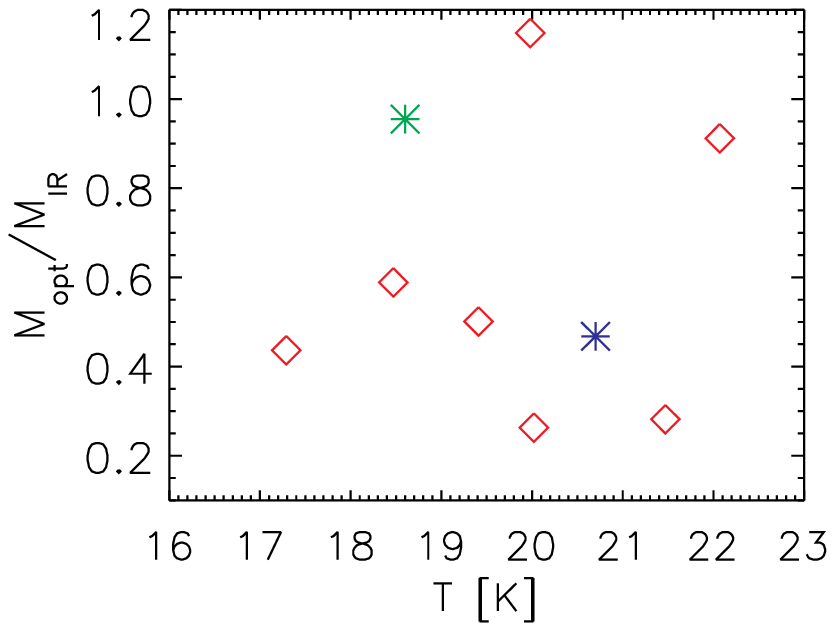} 
  \includegraphics[width=\columnwidth]{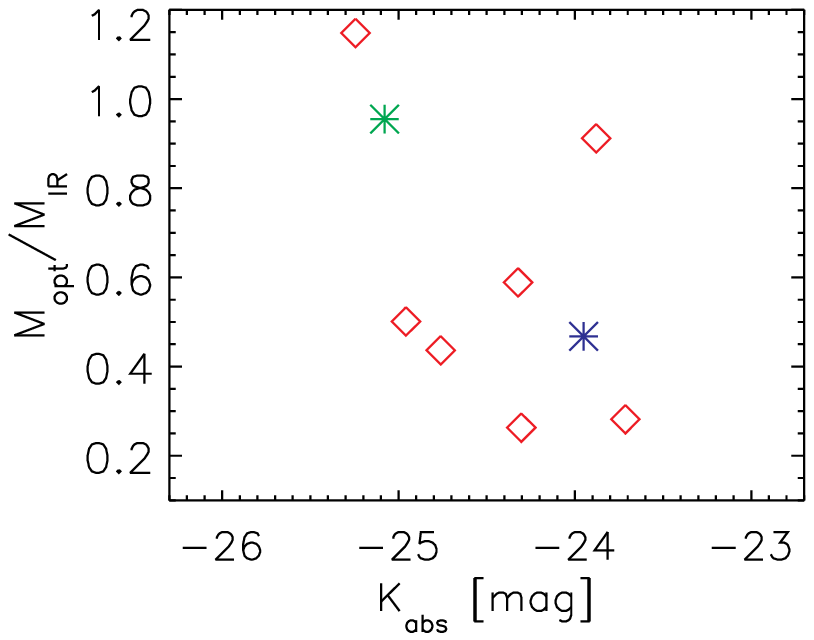} 
  \includegraphics[width=\columnwidth]{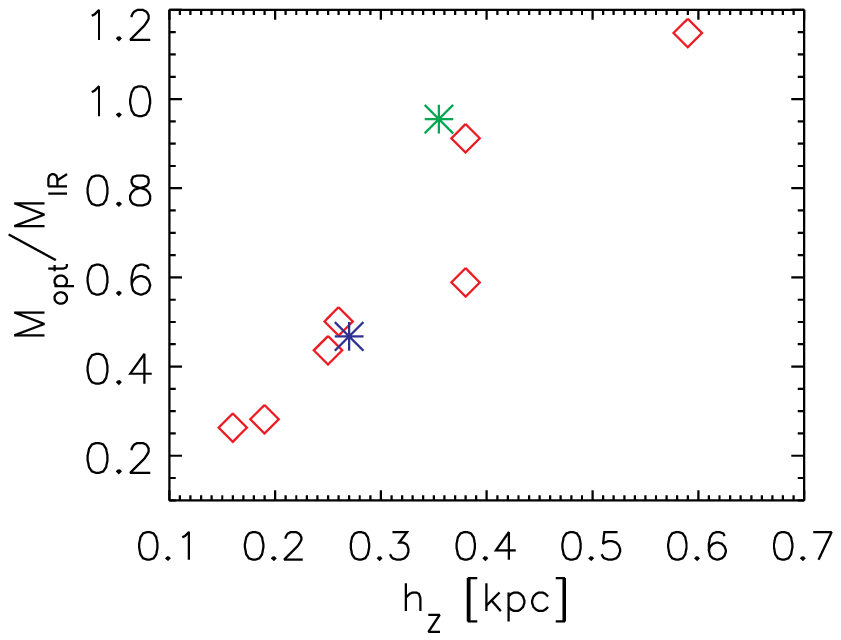} 
  \caption{In these four panels we compare the difference in the
    values of the dust mass as computed from radiative transfer
    modelling of optical data and as computed from black-body fitting
    of \textit{Herschel} datapoints, with various physical quantities: the
    (optically derived) mass of dust, the dust temperature, the
    absolute K-band magnitude and the dust scaleheight. The starred
    points correspond to two other well known edge-on galaxies not
    included in our sample: NGC\,891 (blue) and NGC\,4565 (green).}
\label{DustMassComparison.fig}
\end{figure*}

When we compare the total dust masses as they were determined from
both methods described in the previous subsections, we see a clear
difference between the different mass estimates, with the ones derived
from radiative transfer fits of the optical images being consistently
smaller than the dust masses derived from the far-infrared
emission. Only for NGC\,973, both dust masses are in agreement, while
for all other galaxies the optically determined dust mass
significantly underestimates the FIR dust mass, by a factor of up to
about four. Such a mass discrepancy was previously found in a number
of edge-on spiral galaxies \citep{2000A&A...362..138P,
  2001A&A...372..775M, 2004A&A...425..109A, 2005A&A...437..447D,
  2008A&A...490..461B, 2010A&A...518L..39B, 2012MNRAS.427.2797D}. Our
new observations, which cover also the sub-mm part of the SED and hence
allow a solid determination of the bulk of the cold dust, confirm the
results of these studies.

In order to investigate the underlying mechanism of this discrepancy,
we look for possible relations between the ratio of the optical and
FIR dust masses and a number of global galaxy
properties. Figure~{\ref{DustMassComparison.fig}} plots the
$M_{\text{d}}^{\text{opt}}/M_{\text{d}}^{\text{FIR}}$ as a function of
dust mass, dust temperature, K-band absolute magnitude (used as a
proxy for stellar mass), and dust vertical scaleheight. To increase
the statistics, we include two additional notorious edge-on spiral
galaxies in this figure, i.e.\ the prototypical edge-on spiral
NGC\,891 (blue asterisk) and NGC\,4565, also known as the Needle
Galaxy (green asterisk). To determine the FIR dust mass, we have used
\textit{Herschel} fluxes from \citet{2011A&A...531L..11B} and
\citet{2012MNRAS.427.2797D}, respectively, and a similar fitting
technique was used as for the \textit{HER}OES galaxies. Optical dust
masses were taken from radiative transfer modelling from
\citet{1998A&A...331..894X} and \citet{2004A&A...425..109A}. We have
assumed distances of 9.5 and 16.9 Mpc for NGC\,891 and NGC\, 4565,
respectively.

Looking at Figure~{\ref{DustMassComparison.fig}}, we see no
correlation between the dust mass discrepancy and the dust mass,
stellar mass or dust temperature. Interestingly, we do find a trend
between the dust mass discrepancy and the dust scaleheight as derived
from radiative transfer fits to the optical images. To understand this
trend, we note that the dust mass computed from extinction in optical
images is only sensitive to the smoothly distributed fraction: in
fact, the physical size of the dense, optically thick molecular clouds
often hosting regions of active star formation, is too small for being
detected by optical observations even for such relatively nearby
galaxies, as those of the \textit{HER}OES sample are. The net effect
is that these small clumpy regions do not contribute to the global
optical extinction at all. Their presence can only be revealed from
the thermal emission of the dust they contain, showing up in the FIR
\citep[see e.g.][]{2000A&A...362..138P}. Hence, one of the possible
interpretations of this dust mass discrepancy is that this ratio would
be a measure of the clumpiness of the dust fraction of the ISM in a
galaxy. A correlation between this difference and the scaleheight of
dust would imply that the thinner the dust disc is, the clumpier its
structure is. With only 9 objects being considered, one of which not
lying perfectly on the correlation, we cannot be conclusive on this
point, even though our analysis hints at a real relation, and it is
furthermore supported by a Spearman correlation coefficient of 0.92.

\section{Discussion and conclusions}
\label{Conclusions.sec}

We have presented \textit{Herschel} observations of a sample of
seven edge-on spiral galaxies within the \textit{HER}OES
project. This work is the first one of a series of papers aiming at a
systematic study of the properties of dust, and its relation to the
stellar and gas components, in edge-on spiral galaxies. Here we have
presented FIR and sub-mm data obtained with the \textit{Herschel} Space
Observatory, describing and analysing both the morphology and
the horizontal and vertical distribution of dust.

We have measured the global FIR/sub-mm fluxes of the galaxies in the
PACS and SPIRE bands using aperture photometry. We have compared the
\textit{Herschel} fluxes with \textit{IRAS},
\textit{ISO}, \textit{Akari} and \textit{Planck}
fluxes at similar wavelengths. We find excellent agreement between
\textit{Herschel} on the one side and \textit{IRAS},
\textit{ISO} and \textit{Planck} on the other side. The
only exception is NGC\,5907, where both the \textit{IRAS}
100~$\mu$m\ and the \textit{ISO} 160~$\mu$m\ fluxes
underestimate the PACS fluxes. We argue that this disagreement might
be due to the large extent of this galaxy, which is spatially resolved
even at the course resolution of \textit{IRAS} and
\textit{ISO}. When we compare the PACS 160~$\mu$m\ fluxes with
the \textit{Akari} 160~$\mu$m\ fluxes, we find a strong
inconsistency, with the \textit{Akari} fluxes a factor two
smaller, possibly due to a flux measurement effect.

We have described the resulting FIR morphology.  A double
  exponential disc model for the dust distribution is capable of
  providing a good description of the observed IR and sub-mm profiles,
  especially at 500~$\mu$m, and as long as disc truncation is not
  taken into account.  But at shorter wavelengths, the picture is
more complicated: on the one hand, the occurrence of primary and
several secondary peaks in the horizontal FIR/sub-mm profiles give a
clear indication of morphological structure in the form of arms, rings
or individual star formation complexes. On the other hand, the
underlying horizontal data profiles (taking abstraction of the peaks)
show a more complicated and varying behaviour across the horizontal
span of the galaxies and especially at the PACS wavelengths. Quite
interestingly, the two galaxies likely to host some nuclear activity,
and classified as Seyfert 2 and LINER (NGC\,973 and NGC\,4013
respectively), are those showing the most prominent central peak at
100~$\mu$m, giving a hint of the presence of warmer or more
concentrated dust.  We have checked whether these peaks
  are compatible with a point-like emission, as expected --at these
  resolutions-- for an AGN--like source. We found that the central
  peak is compatible with a gaussian profile emission, imposing upper
  limits on the FWHM of the central source of $\sim$450 and $\sim$120
  pc for NGC\,973 and NGC\,4013 respectively. These values are quite
  high if compared to the physical scales, of the order of few pc,
  found for the dusty tori of local low luminosity AGNs \citep[see
  e.g.][]{2004Natur.429...47J}, whose emission peaks anyway around 30
  to 50~$\mu$m \citep[as IR modelling indicates; see
  e.g.][]{2006MNRAS.366..767F,2012MNRAS.420.2756S}.  Our upper limits
  fit a picture where the intense UV/optical radiation field emitted
  by the central source would heat the dust to slightly large scales,
  but to lower temperatures. This hypothesis has so far never been
  tested, but it is in principle verifiable by means of radiative
  transfer models.

By fitting an exponential model to the vertical profile
  of the \textit{Herschel} images, we investigate whether we can
  detect vertically extended dust in edge-on galaxies. Evidence for
  extra-planar dust has been found before, either by means of
  extinction features in high-resolution optical images
  \citep{1999AJ....117.2077H, 2000A&A...356..795A,
    2004AJ....128..662T} or through warm dust or PAH emission at
  mid-infrared wavelength \citep{2006A&A...445..123I,
    2007A&A...474..461I, 2007A&A...471L...1K,
    2009MNRAS.395...97W}. Three out of the seven galaxies show
  signatures of extended vertical emission at 100 and 160~$\mu$m at
  the 5$\sigma$ level. For two of these three galaxies (NGC\,4217 and
  NGC\,5907), this vertically extended emission is most probably due
  to projection effects as a result of deviations from an exactly
  edge-on orientation. For the remaining galaxy, NGC\,4013, the
  FIR/sub-mm emission seems truly resolved, and the inferred
  scaleheights are in agreement with the scaleheight
  independently derived from radiative transfer modelling of the optical
  images by \citet{2007A&A...471..765B}. We find a hint of an
  increase in the scaleheight with FIR wavelength; rather than
  interpreting this as evidence that the dust temperature decreases
  with increasing distance above the plane of the galaxy, we argue
  that this is probably due to the limited and gradually worsening
  resolution of the {\em{Herschel}} images for increasing
  wavelengths.

Finally, total dust masses inferred from the optical extinction
through radiative transfer models were compared with those determined
from modified black-body fits to the FIR fluxes, and correlations with
other physical and geometrical properties of the galaxies were
searched for. While we do acknowledge that our sample --which we
extended with two other well-known edge-on galaxies
galaxies for this particular analysis-- is quite limited to provide
strong evidence, we found a hint of a correlation between the
discrepancy between both dust masses and the vertical scaleheight of
the dust: larger differences are found in galaxies with smaller
scaleheights.  If the discrepancy between the dust mass as
  derived from optical extinction and that calculated from IR emission
  is interpreted as a measure of the clumpiness of the ISM, this
would be consistent with a picture where dust which is more
``compressed'' into the disc, would tend to be gathered in clumps more
strongly, as opposed to a smoother, continuous
distribution.  While we are not aware of any study
  addressing this issue, this correlation will be tested by means of
  radiative transfer models. By exploiting a state-of-the-art
  radiative transfer code coupled with a robust fitting algorithm, we
  should be able to at least confirm the presence of this trend. In
  fact, following the approach already used by \citet{2000A&A...362..138P, 2000A&A...359...65B} and more recently
  \citet{2012MNRAS.427.2797D}, the FIR luminosity deficit always
  observed in models can be accounted for by including in the model,
  a posteriori, a dust emission component originating from very
  compact regions, hence invisible in optical extinction maps. The
  relative amount of this component with respect to the diffuse dust
  can then be compared to the dust scaleheight derived from the model
  itself and this, in turn, checked against our findings.

\begin{acknowledgements}

  JV, MB, FA, GDG, GG and SV acknowledge the support of the Flemish
  Fund for Scientific Research (FWO-Vlaanderen). JF, MB and JADLB are
  grateful for the support from the Belgian Science Policy Office
  (BELSPO).

  PACS has been developed by a consortium of institutes led by MPE
  (Germany) and including UVIE (Austria); KU Leuven, CSL, IMEC
  (Belgium); CEA, LAM (France); MPIA (Germany); INAFIFSI/ OAA/OAP/OAT,
  LENS, SISSA (Italy); IAC (Spain). This development has been
  supported by the funding agencies BMVIT (Austria), ESA-PRODEX
  (Belgium), CEA/CNES (France), DLR
  (Germany), ASI/INAF (Italy), and CICYT/MCYT (Spain).

  SPIRE has been developed by a consortium of institutes led by
  Cardiff University (UK) and including Univ. Lethbridge (Canada);
  NAOC (China); CEA, LAM (France); IFSI, Univ. Padua (Italy); IAC
  (Spain); Stockholm Observatory (Sweden); Imperial College London,
  RAL, UCL-MSSL, UKATC, Univ. Sussex (UK); and Caltech, JPL, NHSC,
  Univ. Colorado (USA). This development has been supported by
  national funding agencies: CSA (Canada); NAOC (China); CEA, CNES,
  CNRS (France); ASI (Italy); MCINN (Spain); Stockholm Observatory
  (Sweden); STFC (UK); and NASA (USA).

  HSpot and HIPE are joint developments by the \textit{Herschel}
  Science Ground Segment Consortium, consisting of ESA, the NASA
  \textit{Herschel} Science Center, and the HIFI, PACS and SPIRE
  consortia.

  This research has made use of NASA's Astrophysics Data System, and
  of the NASA/IPAC Extragalactic Database (NED) which is operated by
  the Jet Propulsion Laboratory, California Institute of Technology,
  under contract with the National Aeronautics and Space
  Administration.
  
  We wish to thank the anonymous referee for the comments and suggestions, which helped us to improve this work.
\end{acknowledgements}

\bibliographystyle{aa} 
\bibliography{HEROES}

\appendix

\section{The effect of the inclination on the
    vertical structure} 

  In this Appendix we investigate how deviations from an exactly
  edge-on orientation affect the apparent vertical structure of
  edge-on galaxies in the FIR/sub-mm maps.  Consider first a
  double-exponential disc, described by the three-dimensional density
  distribution~(\ref{ded}) and the dust surface density
  distribution~(\ref{ded-Sigma}) when viewed exactly edge-on.  When we
  collapse the surface density distribution in the horizontal
  direction and normalise the resulting expression, we recover a
  simple exponential function as the vertical profile 
\begin{equation}
  \Sigma_{\text{ver}}(y)
  =
  \frac{1}{M_{\text{d}}}
  \int_{-\infty}^\infty \Sigma(x,y)\,{\text{d}}x
  =
  \frac{1}{2h_z}\exp\left(-\frac{|y|}{h_z}\right)
\label{Sigmacexp}
\end{equation}
Now assume a system in which the dust is distributed in an infinitely
thin, exponential disc, i.e.\ with a density distribution like
Eq.~(\ref{ded}) with $h_z\rightarrow0$,
\begin{equation}
  \rho(R,z)
  =
  \frac{M_{\text{d}}}{2\pi\,h_R}
  \exp\left(-\frac{R}{h_R}\right)
  \delta(z)
\end{equation}
When this disc has an inclination $i$, it has as mass
surface density distribution projected on the sky
\begin{equation}
  \Sigma(x,y)
  =
  \frac{M}{2\pi\,h_R^2 \cos i}
  \exp\left(-\frac{1}{h_R}\sqrt{x^2+\frac{y^2}{\cos^2 i}}\right)
\label{Sigmaxy}
\end{equation}
When we collapse this surface density profile in the
horizontal direction and normalise, we find
\begin{equation}
  \Sigma_{\text{ver}}(y)
  =
  \frac{1}{\pi\,h_R \cos i}
  \left(\frac{|y|}{h_R\cos i}\right)\,
  K_1\left(  \frac{|y|}{h_R\cos i}\right)
\label{Sigmac}
\end{equation}
When an infinitesimally thin exponential disc is not exactly edge-on,
it will hence give rise to a vertical profile that is extended, which
could be falsely interpreted as an intrinsical vertical distribution
of an exactly edge-on spiral galaxy. To determine the
importance of this inclination effect, we estimate the exponential
scaleheight $h_z$ that would correspond to the vertical profile
(\ref{Sigmac}), by requiring the FWHM of both distributions are
equal. For the expression (\ref{Sigmac}) we find 
\begin{equation}
  {\text{FWHM}}
  =
  2.5143\,h_R\cos i
\end{equation}
whereas the exponential profile~(\ref{Sigmacexp}) gives
\begin{equation}
  {\text{FWHM}}
  =
  2\ln 2\,h_z = 1.3863\,h_z
\end{equation}
and therefore
\begin{equation}
  h_z 
  =
  1.8137\,h_R\cos i
\end{equation}
The values we obtain using this formula for the seven {\em{HER}}OES
galaxies are listed in the last column of Table~{\ref{tab:hz}}.

\end{document}